\documentclass[useAMS,usenatbib,twocolumn]{mn2e}
\usepackage{amsmath,amssymb}
\usepackage{natbib}
\usepackage[dvipdfmx]{graphicx}
\usepackage{times}
\usepackage{rotate}
\usepackage{color,url}
\newcommand{\simgt}{\lower.5ex\hbox{$\; \buildrel > \over \sim \;$}}
\newcommand{\simlt}{\lower.5ex\hbox{$\; \buildrel < \over \sim \;$}}

\newcommand{\redmapper}{redMaPPer}

\newcommand{\eg}{{\it e.g.}}

\begin{document}
  

 \title[Luminous Red Galaxies in Clusters]{Luminous Red Galaxies in
   Clusters: Central Occupation, Spatial Distributions, and Mis-centering}




 \author[Hoshino et al.]
        {Hanako Hoshino$^1$, 
         Alexie Leauthaud$^2$, 
         Claire Lackner$^2$, 
         Chiaki Hikage$^3$,
         Eduardo Rozo$^4$, \newauthor 
         Eli Rykoff$^5$, 
         Rachel Mandelbaum$^6$,
         Surhud More$^2$, 
         Anupreeta More$^2$, 
         Shun Saito$^2$, \newauthor 
         Benedetta Vulcani$^2$\\
$^1$ Department of Physics and Astrophysics, Nagoya University, Aichi 464-8602, Japan \\
$^2$ Kavli IPMU (WPI), UTIAS, The University of Tokyo, Kashiwa, Chiba 277-8583, Japan\\
$^3$ Kobayashi-Maskawa Institute, Nagoya University, Nagoya 464-8602, Japan\\
$^4$ Department of Physics, University of Arizona, 1118 E. Fourth St., Tucson, AZ 85721, USA \\
$^5$ SLAC National Accelerator Laboratory, Menlo Park, CA 94025, USA \\
$^6$ McWilliams Center for Cosmology, Department of Physics, Carnegie Mellon University, 5000 Forbes Ave., Pittsburgh, PA 15213, USA}

\maketitle
\label{firstpage}

\begin{abstract} 
  Luminous Red Galaxies (LRG) from the Sloan Digital Sky Survey are
  considered among the best understood samples of galaxies, and they
  are employed in a broad range of cosmological studies. Because they
  form a relatively homogeneous population, with high stellar masses
  and red colors, they are expected to occupy halos in a relatively
  simple way. In this paper, we study how LRGs occupy massive halos
  via direct counts in clusters and we reveal several unexpected
  trends suggesting that the connection between LRGs and dark matter
  halos may not be straightforward. Using the \redmapper{} cluster
  catalog, we derive the central occupation of LRGs as a function
  richness, $N_\text{cen}(\lambda)$. Assuming no correlation between
  cluster mass and central galaxy luminosity at fixed richness, we
  show that clusters contain a significantly lower fraction of central
  LRGs than predicted from the two-point correlation function. At halo
  masses of $10^{14.5} M_\odot$, we find $N_{\text{cen}}=0.73$,
  compared to $N_{\text{cen}}$ of $0.89$ from correlation studies. Our
  central occupation function for LRGs converges to $0.95$ at large halo
  masses. A strong anti-correlation between central luminosity and
  cluster mass at fixed richness is required to reconcile our
  results with those based on clustering studies. We also derive
  $P_\text{BNC}$, the probability that the brightest cluster member is
  not the central galaxy. We find $P_\text{BNC}\approx 20-30\%$ which
  is a factor of $\sim2$ lower than the value found by
  \citet{Skibba:2011}. Finally, we study the radial offsets of bright
  non-central LRGs from cluster centers and show that bright
  non-central LRGs follow a different radial distribution compared to
  red cluster members, which follow a Navarro-Frank-White
  profile. This work demonstrates that even the most massive clusters
  do not always have an LRG at the center, and that the brightest
  galaxy in a cluster is not always the central galaxy.
\end{abstract}

\begin{keywords}
   galaxies: clusters: general
\end{keywords}
 

 
 


\section{Introduction}
\label{sec:intro}

Luminous Red Galaxies (LRGs) are early-type massive galaxies
consisting mainly of old stars with little ongoing star
formation. Because LRGs are very luminous and can be detected to $z
\sim 0.5$, they are commonly used for studies of large scale structure. The
Sloan Digital Sky Survey \citep[SDSS,][]{York:2000} targeted a large
sample of LRGs out to $z \sim 0.5$ using color and magnitude cuts
described in \citet{Eisenstein:2001}. LRGs from SDSS have been used
for a wide variety of purposes, such as the detection of baryon
acoustic oscillations \citep[\eg,][]{Eisenstein:2005,Kazin:2010}, clustering
studies \citep[\eg,][]{Zehavi:2005,Reid:2009,Padmanabhan:2009,Parejko:2013},
weak lensing and cross-correlation studies 
\citep[\eg,][]{Mandelbaum:2006,Hikage:2013}, and redshift space distortion
(RSD) studies
\citep{Cabre:2009aa,Reid:2009aa,Samushia:2012aa,Hikage:2013aa}.

All of the studies described above require an understanding of how
LRGs trace dark matter halos, and, therefore, large scale
structure. One important aspect of how LRGs trace dark matter is how
often they coincide with the centers of dark matter halos. The
identification of central galaxies in cluster samples is critical for
a variety of studies, including the determination of cluster halo
masses with weak gravitational lensing \citep{George:2012}, and RSD
studies \citep{Hikage:2013aa}. Because LRGs are luminous and trace
massive overdensities, it is often assumed that central galaxies are
also LRGs. Indeed, in most galaxy clusters with at least one LRG, the
Brightest Cluster Galaxy (BCG) is an LRG. Furthermore, it is usually
assumed that BCGs are {\em central} galaxies
\citep{van-den-Bosch:2004,Weinmann:2006,Budzynski:2012}. However,
recent results suggest that the central galaxy is not the BCG in all clusters
\citep{van-den-Bosch:2005aa,Coziol:2009aa,Sanderson:2009aa,
  Skibba:2011,Einasto:2011,Einasto:2012,Hikage:2013,Sehgal:2013,Lauer:2014}. 
For example, by analyzing differences
between the velocities and positions of BCGs relative to other cluster
members, \citet{Skibba:2011} found that 40\% of BCGs may in fact be
satellite galaxies.

Knowing how LRGs occupy dark matter halos, and especially whether or
not the most dominant LRGs coincides with cluster centers, is of
particular importance for RSD studies. \citet{Hikage:2013aa} found
that satellite LRGs have a significant effect on the higher order
multipoles of the LRG power spectrum. This is simply due to the fact
that satellite galaxies have large peculiar velocities that cause
structures to be smeared out along the line-of-sight (the
``Finger-of-God'' effect). Hence, understanding if LRGs trace halo
centers and how often LRGs are satellites is an important
ingredient for large scale structure studies with LRG samples.

Halo Occupation Distribution models \citep[HOD:][]{Seljak:2000,
  Peacock:2000, Scoccimarro:2001, Berlind:2002} are the most popular
framework for describing how LRGs populate dark matter halos. The HOD
model relies on the choice of a functional form for the probability
that a halo of mass $M_\text{h}$ contains $N$ objects of a particular
type ($P(N|M_{\rm h})$). This probability is typically divided into a
contribution from central and from satellite galaxies. The mean
probability for each component is the central occupation function,
$N_\text{cen}$, and the satellite occupation function,
$N_\text{sat}$, respectively. At group and cluster scales $(M_\text{h} > 10^{14}
M_\odot)$, the central occupation function of LRGs is often assumed to
converge to unity. If this assumption is correct, then the central
galaxies of massive halos should all be LRGs. One of the primary goals
of this paper is to test this assumption.

HOD models for LRGs are typically constrained by measurements of the
LRG two-point correlation function \citep[\eg,][]{Zheng:2005, 
Masjedi:2006aa} or by using the counts-in-cylinders method developed by 
\citet{Reid:2009}. However, a complementary way to measure the HOD of 
LRGs is to directly count LRGs in clusters as a function of halo mass. 
For example, \citet{Ho:2009} studied the HOD of LRGs using a sample of 
47 clusters from the ROentgen SATellite (ROSAT) X-ray survey. The main 
caveat to this direct approach is that it relies on the availability of 
a trustworthy cluster catalog with known centers and halo masses. Early 
cluster catalogs from SDSS such as MaxBCG \citep{Koester:2007} are known 
to have problems with centering. \citet{Johnston:2007aa}
estimate that the mis-identification of centrals in the MaxBCG catalog
is roughly $\sim 30\%$. Recently, however, much progress has been made
in understanding how to improve centering algorithms for cluster
finders \citep{Rykoff:2014, Rozo:2014}.

The goal of this paper is to compare the central occupation function
of LRGs as inferred from galaxy correlation measurements to direct counts of
central LRGs in clusters. We use the redMaPPer cluster catalog
\citep{Rykoff:2014, Rozo:2014}, which includes carefully selected
central galaxies based on luminosity, color and local galaxy
density. One of the advantages of redMaPPer over previous cluster
finders is that it adopts a probabilistic approach to cluster
centering. This probabilistic approach is especially useful when, for
example, two clusters are merging and when the central galaxy is not
obvious. We caution that while the comparison of X-ray to optical
centers in galaxy clusters  so far suggests that
the \redmapper{} centering probabilities are accurate
\citep{Rozo:2014}, the samples used in these comparisons are still
small, and rely heavily on X-ray selected sub-samples of clusters.
More detailed investigation into the \redmapper{} centering
probabilities is still warranted, but for the purposes of this study
we will assume the centering probabilities from the redMaPPer
algorithm are correct. In Paper II, we perform additional tests on the
redMaPPer centering probabilities by using weak gravitational lensing
and cross-correlations. Preliminary results from this work indicate
that the redMaPPer centroids are, on average, better indicators of
halo centers than BCGs, adding confidence to
the redMaPPer centering probabilities. 

This paper is organized as follows. In $\S$\ref{sec:data} we briefly
describe the redMaPPer cluster catalog and our photometric SDSS LRG
selection. In $\S$\ref{sec:results}, we present the central occupation
function of LRGs as inferred from the redMaPPer cluster catalog, and 
compare four different
definitions of the central galaxy. Our
main results are presented in $\S$\ref{sec:results}. Finally, we
discuss the results and draw our conclusions in
$\S$\ref{sec:discussion} and $\S$\ref{sec:conclusions}.

Throughout this paper, we adopt a flat $\Lambda$CDM cosmology with 
$h=0.7$, $\Omega_m = 0.25$, and $\sigma_8=0.8$. This cosmology is 
chosen for consistency with the redMaPPer richness-halo mass 
calibration in \citet{Rykoff:2012}. All distances are expressed in 
physical ${\rm Mpc}$ units. Halo mass is defined as 
$M_{\rm 200b}\equiv M(<R_{\rm 200b})=200\bar{\rho} 
\frac{4}{3}\pi R_{\rm 200b}^3$, where $R_{\rm 200b}$ is the radius at 
which the mean interior density is equal to 200 times the mean matter 
density ($\bar{\rho}$). All magnitudes are given on the AB system.


\section{Data}
\label{sec:data}

\subsection{RedMaPPer Cluster Catalog}
\label{sec:redmapper-catalog}

The red-sequence Matched-filter Probabilistic Percolation 
(redMaPPer) algorithm is a red-sequence based cluster-finder 
optimized for identifying clusters in large multi-wavelength 
surveys. The details of the algorithm can be found in 
\citet{Rykoff:2014} and \citet{Rozo:2014}. Comparisons between 
redMaPPer and other cluster finding algorithms are performed 
in \citet{Rozo:2014,Rozo:2014a}. The key elements of the 
redMaPPer algorithm are briefly summarized here. Using galaxies 
with spectroscopic redshifts, redMaPPer first constructs a 
red-sequence model as a function of luminosity and redshift. 
Once the red-sequence model is calibrated, redMaPPer then uses 
an iterative method to identify clusters assuming simple radial
and luminosity filters. The redMaPPer algorithm is fully
probabilistic. All potential cluster members are assigned a
probability, $p_\text{mem}$, of being a red cluster member. 
Cluster membership probabilities are tested using spectroscopic 
redshifts and are accurate to 1\% \citep{Rozo:2014b}. Each 
cluster is assigned a photometric redshift, denoted $z_{\lambda}$, 
that is estimated from high-probability cluster members. Finally, 
redMaPPer estimates cluster richness (denoted $\lambda$) as the 
sum of membership probabilities over galaxies:
\begin{equation}
   \lambda = \sum_i p_{\text{free},i} \ p_{\text{mem},i}
\end{equation}
where $p_\text{free}$ represents the probability that a galaxy 
is not a member of another cluster (in general, $p_\text{free} 
\approx 1$). The sum is performed over all galaxies within a 
characteristic cutoff radius, $R_\lambda$, which is a richness-dependent 
aperture that minimizes scatter in the mass-richness relation.

A key feature of the redMaPPer algorithm is that cluster centering 
is also probabilistic. For each cluster, redMaPPer provides a list 
of the five most likely central galaxies. Each of these five 
central galaxy candidates is assigned a probability, $P_\text{cen}$, 
of being the cluster center. Centering probabilities are defined 
using three filters: a luminosity filter ($\phi_\text{cen}$), a 
photometric redshift filter ($G_\text{cen}$), and a local galaxy 
density filter ($f_\text{cen}$). The product of these three terms 
determines the overall centering filter:
\begin{equation}
   u_\text{cen} = \phi_\text{cen} (m_i | z_\lambda,\lambda)
                 G_\text{cen} (z_\text{red})
                 f_\text{cen} (w | z_\lambda,\lambda),
\label{eq:central-filter}
\end{equation}
Here, $m_i$ is the $i$-band apparent magnitude, $z_\text{red}$ is the
redshift of an individual galaxy (inferred from the color), $w$ is the
local galaxy density. An important point to note is that the redshift
filter $G_\text{cen}$ is broader than the cluster red-sequence filter
and allows galaxies with slightly offset colors, and, therefore, lower
$p_\text{mem}$ values, to be considered central galaxy
candidates. This allows for the possibility of residual amounts of
star-formation, and, therefore, bluer colors, in centrals. For each cluster, centering probabilities
are normalized as:
\begin{equation}
   \sum_{i=1}^5 P_{\text{cen},i} = 1
\end{equation}

One consequence of Equation (\ref{eq:central-filter}) is that the 
brightest cluster member is not necessarily the galaxy with the 
highest value of $P_\text{cen}$. This is because, in addition to 
luminosity, the central galaxy selection also considers local 
galaxy density via the $f_\text{cen}$ filter. Hence, in some cases, 
a less luminous galaxy may have a higher centering probability 
because it is spatially coincident with the dense cluster core.

We use the redMaPPer v5.10 cluster catalog \citep{Rozo:2014b} based on
the SDSS 8th Data Release \citep[DR8;][]{Aihara:2011}. We select
clusters in the range $\lambda > 20$ and $0.16 < z < 0.33$. The lower
redshift limit, $z_\text{min} = 0.16$, is set to avoid low redshifts
where the LRG color cuts begin to select a fainter population of
galaxies \citep{Eisenstein:2001}. The upper redshift limit,
$z_\text{max} = 0.33$, is set so that the resulting sample of clusters
is volume limited. In addition, we select clusters for which fewer
than 20\% of the cluster is masked (\verb+mask_frac+ $<$0.2).
Altogether, there are 7,730 clusters in this redshift and richness
range.

Our SDSS DR8 redMaPPer cluster catalog is limited to $i<21.0$. This is
roughly one magnitude deeper than LRG samples (see Section
\ref{sec:classical-LRGs}), ensuring that our cluster membership
selection is complete for LRG-type galaxies.

\subsection{Luminous Red Galaxies}
\label{sec:lrg}

LRGs are intrinsically red and bright galaxies selected from SDSS
\citep{Eisenstein:2001}. LRG samples extend fainter and farther than
the SDSS main galaxy sample \citep{Strauss:2002} and consist of mainly
passively-evolving, massive, early-type galaxies with redshifts in the
range $0.16<z<0.7$ (though we restrict the redshift range to
$0.16<z<0.33$ here). In this paper, we will distinguish ``classical
LRG'' samples, which were targeted as part of the SDSS-I and SDSS-II
programs \citep{Eisenstein:2001}, from LRGs that were targeted as
part of the SDSS-III Baryon Oscillation Spectroscopic Survey (BOSS)
program \citep{Dawson:2013}. We will refer to LRGs selected in SDSS-I
and II as ``Classical LRGs''.  We will use the term ``LOWZ'' to refer
to low redshift ($z<0.4$) LRGs from the SDSS-III BOSS program.

Fiber collisions occur because two fibers on the same SDSS
spectroscopic plug plate (referred to as a ``tile'') cannot be placed
closer than 55 arcsec (SDSS-I/II) or 62 arcsec (SDSS-III). Large
clusters may contain several LRGs and are hence more likely to be
affected by fiber collisions than smaller systems, which have
only one LRG. As a result, in the redMaPPer catalog, 23\% of LRG
cluster members lack a spectroscopic redshift. In order to circumvent
this issue, we will use photometrically-defined LRG samples.  We use
photometric data from SDSS DR8 for both the classical LRG and LOWZ
selections.

In this paper, we consider the following three LRG samples:

\begin{enumerate}
\item DR8PhotLRG : photometrically defined ``classical'' LRGs 
selected from DR8 photometry.
\vspace{0.5\baselineskip}

\item DR8PhotLRG-MgCut : same as above but with an additional $g$-band
  absolute magnitude cut (see Section \ref{sec:classical-LRGs}). We
  use this sample to compare with previous results derived from
  clustering studies.  \vspace{0.5\baselineskip}

\item PhotLOWZ : photometrically defined BOSS LOWZ sample selected 
from DR8 photometry.
\end{enumerate}

We base our LRG selections on DR8 photometry but we note that sub-sets
of the LRG samples were targeted based on photometry from earlier data
releases. In Appendix A, we compare our samples based on DR8
photometry to LRG samples based on DR7 photometry and to the original
spectroscopic target catalogs. In addition to covering a larger area
than DR7, the DR8 photometry uses a new sky subtraction algorithm,
affecting mainly bright galaxies ($r < 16$) \citep{Abazajian:2009},
and a new calibration \citep{Padmanabhan:2008}. While the total
numbers of LRGs selected using DR7 and DR8 photometry are the same,
the actual galaxy samples differ at the $\sim 5\%$ level. This is due
to scatter around the color and magnitude cuts imposed by the LRG
selections below. As we show in the Appendix A, the differences in
photometry for LRGs are small (clipped RMS $\delta m_r = 0.07$) and
there are no systematic offsets in magnitudes or color. The lack of
systematic offsets suggests that the improved sky subtraction in DR8
has little effect on LRGs. This is expected since
most LRGs are fainter than $r<16$, where the effects of improved sky
subtraction are most noticeable. Therefore, while the DR7-selected and
DR8-selected LRG samples are not identical, they are comparable.

\subsection{Classical LRGs and DR8PhotLRG Cluster Member Selection}
\label{sec:classical-LRGs}

In this section we explain how we construct our sample of classical
LRG cluster members, the ``DR8PhotLRG'' sample. As stated in Section
\ref{sec:redmapper-catalog}, we only consider cluster members with
cluster redshifts between $0.16 < z < 0.33$. This redshift range is
similar to the range considered for a variety of LRG studies:
\citet{Zehavi:2005} use $0.16<z<0.44$, \citet{Reid:2009} use
$0.16<z<0.36$, and \citet{Kazin:2010} use $0.16<z<0.36$.

Classical LRGs are selected using a series of color-color and
luminosity cuts in SDSS. Because the 4000\AA \ break moves from 
$g$-band to $r$-band at $z \sim 0.4$, for classical LRGs there 
are two separate selections at redshifts above and below $z \sim
0.4$. The LRG selection at $z \lesssim 0.4$ is called ``Cut I''. 
Given our redshift range, in this paper, we only consider 
``Cut I'' LRGs. All colors are defined using model 
magnitudes\footnote{Note that model magnitude were improved in 
SDSS DR2 and later data releases. This caused changes in galaxy 
colors $(g_\text{mod}-r_\text{mod})$ and $(r_\text{mod} - 
i_\text{mod})$ of about 0.005 magnitude. The DR1 LRG selection 
criteria originally specified by \citet{Eisenstein:2001} 
would lead to an increase in the LRG number density by about 10\% 
using DR2 photometry. To account for this, the LRG selection was 
slightly modified.} and all quantities are corrected for galactic
extinction \citep{Schlegel:1998}. 

We query the SDSS {\tt PhotoPrimary} table (see Appendix B for the 
full query) for all cluster members to obtain the photometric 
quantities necessary to apply the LRG ``Cut I'' selection. 
Hereafter, the subscript ``mod'' denotes model magnitudes, which 
are derived by adopting the better fitting luminosity profile 
between a de~Vaucouleurs and an exponential luminosity profile in 
the $r$-band.  
The subscript ``cmod'' denotes composite model magnitudes, which 
are calculated from the best-fitting linear combination of a 
de~Vaucouleurs and an exponential luminosity profile. For 
classical LRGs, luminosity cuts are defined using $r_\text{Petro}$, 
the Petrosian $r$-band magnitude. Two ancillary colors, $c_\perp$ 
and $c_\|$, are defined to align perpendicular and parallel to the 
locus of LRG galaxies in the $(g_\text{mod}-r_\text{mod}, 
r_\text{mod}-i_\text{mod})$ plane:
\begin{align}
  c_\perp &= (r_\text{mod}-i_\text{mod}) 
              - (g_\text{mod}-r_\text{mod})/4 - 0.177, \\
  c_\| &= 0.7 (g_\text{mod}-r_\text{mod}) 
            + 1.2 [(r_\text{mod}-i_\text{mod}) - 0.177].
\end{align}
Galaxies at $z \lesssim 0.4$ are distributed close to the linear 
locus $c_\perp = 0$, and $c_\|$ denotes where galaxies fall along 
this locus. Galaxies at higher redshifts have larger $c_\|$. 

In this paper, we focus only on LRG samples at $z<0.4$ which are
selected via ``Cut I'', as defined on the DR8 target selection webpage\footnote{http://www.sdss3.org/dr8/algorithms/target\_selection.php}:
\begin{align}
   \label{eq:lrg-eq1}
   r_\text{Petro} &< 13.116 + c_\|/0.3, \\
   \label{eq:lrg-eq2}
   r_\text{Petro} &< 19.2, \\
   \label{eq:lrg-eq3}
   | c_\perp | &< 0.2, \\
   \label{eq:lrg-eq4}
   \mu_{50} &< 24.2 \ \mathrm{mag \ arcsec^{-2}}, \\
   \label{eq:lrg-eq5}
   r_\text{psf} - r_\text{mod} &\geq 0.24,
\end{align}
where $\mu_{50}$ is the average surface brightness inside a Petrosian
half light radius, and $r_\text{psf}$ is the point spread function
(PSF) magnitude. Equation (\ref{eq:lrg-eq1}) ensures that the LRG
sample has a roughly constant absolute magnitude limit between
$0.16<z<0.4$. Equation \eqref{eq:lrg-eq2} is set by the $S/N$
requirement for spectroscopic redshift measurements using the SDSS
telescope.  The $c_\perp$ cut limits the redshift range to $z \lesssim
0.4$. The $\mu_{50}$ cut removes very low surface brightness objects,
which may have odd colors.  The final cut is a star-galaxy
separator. As shown in Table \ref{tb:LRG-cut}, Equations
(\ref{eq:lrg-eq1}) and (\ref{eq:lrg-eq2}) have the most impact in
defining the LRG cluster member sample. However, equation
(\ref{eq:lrg-eq2}) only has a small impact on the LRG central sample
simply because centrals are typically brighter than this limit.

\begin{table}
\centering
\caption{Percentage of cluster members and central galaxies that fail
  to pass the LRG selection cuts given by Equations \eqref{eq:lrg-eq1}
  - \eqref{eq:lrg-eq5}. For example, among all 507,874 cluster member
  galaxies, 492,949 (97.1\%) do not satisfy Equation
  \eqref{eq:lrg-eq1}, and 390,840 (77.0\%) do not satisfy Equation
  \eqref{eq:lrg-eq2}.  Each cut is applied separately in this
  Table. Applying all of these cuts 12,001 cluster members (2.36\%)
  and 5,823 central galaxies (75.3\%) are classified as LRGs.}

\begin{tabular}{lccccc} \\ \hline
{\bf LRG cut}  &  Eq. \eqref{eq:lrg-eq1}  &  Eq. \eqref{eq:lrg-eq2}  
   &  Eq. \eqref{eq:lrg-eq3}  &  Eq. \eqref{eq:lrg-eq4}  
   &  Eq. \eqref{eq:lrg-eq5}  \\ \hline
\shortstack{All cluster \\ members}
   &  97.1  &  77.0  &  5.37  &  1.65  &  2.04 \\ \hline
\shortstack{Most likely \\ central galaxies}
   &  24.4  &  0.776  &  0.285  &  0.181  &  0.0129 \\ \hline
\end{tabular}

\label{tb:LRG-cut}
\end{table}

\begin{figure*}
\centering
\includegraphics[width=16cm,clip]{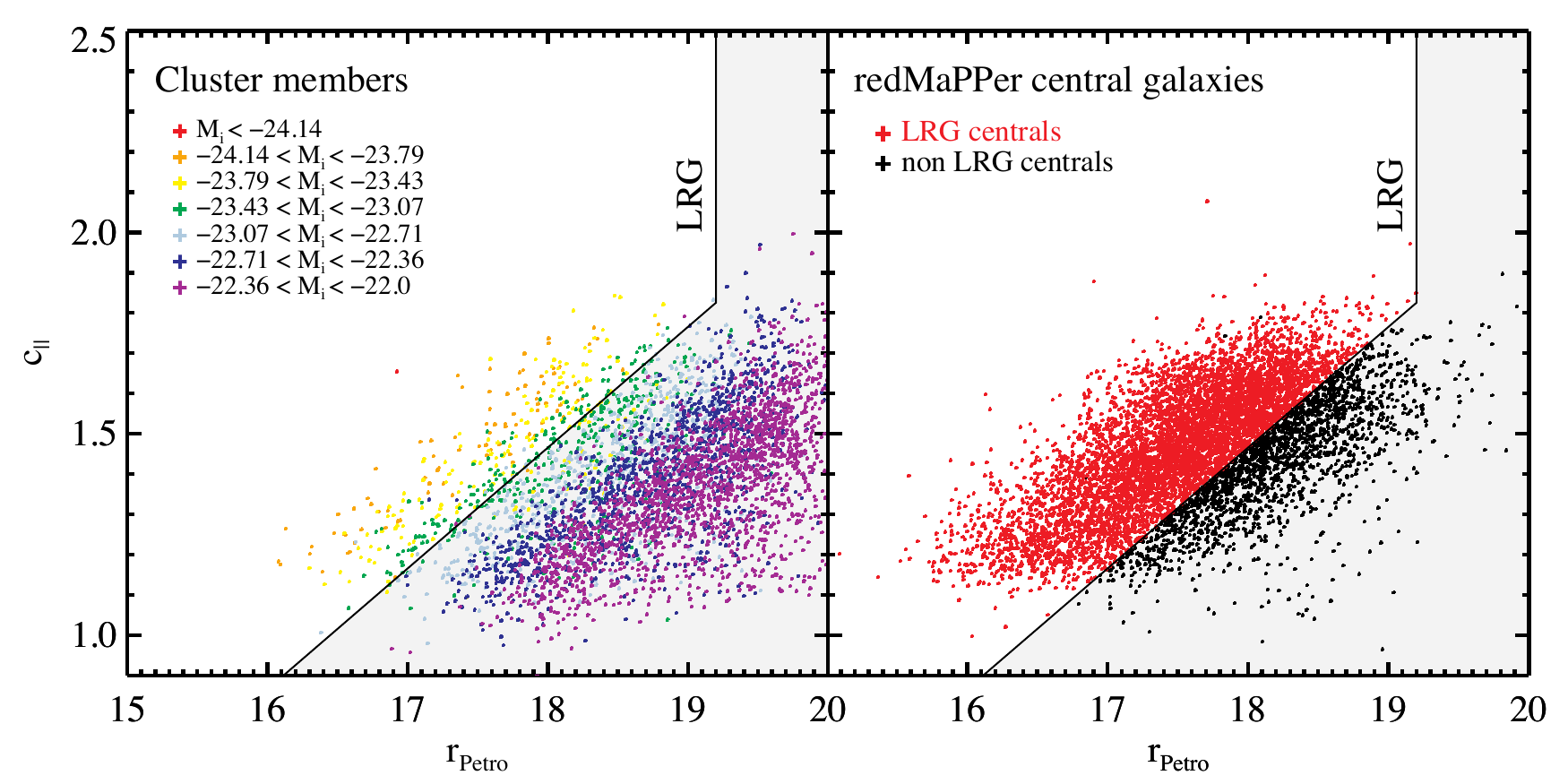}

\caption{{\bf Left panel}: $c_\|$ as a function of Petrosian $r$-band
  magnitude, $r_\text{Petro}$, for cluster members with $0.16<z<0.33$,
  $M_i<-22$ and $\lambda>20$ (data points are randomly down-sampled
  for visual clarity).  Cluster members are color coded by their
  $i$-band absolute magnitude, $M_i$. The classical LRG cuts select
  cluster members brighter than $M_i \sim -23.2$. Among all cluster
  members, about 2\% are classified as LRGs. {\bf Right panel}: $c_\|$
  versus $r_\text{petro}$ for the {\it most likely} central galaxies 
  chosen by redMaPPer. Red points
  show redMaPPer central galaxies that are LRGs while black points
  show centrals that fail the LRG selection. Most of the central
  galaxies in our sample that fail to pass the LRG cuts are
  intrinsically fainter than $M_i \sim -23.2$ (also see Figure
  \ref{fig:central-imag}). Roughly 25\% of central galaxies in our
  cluster sample fail the LRG cuts. Central galaxies are primarily rejected from the LRG sample by Equation (\ref{eq:lrg-eq1}).}
\label{fig:sdss-lrg-cuts}
\end{figure*}

With the extinction-corrected composite magnitudes, we calculate
absolute magnitudes $M_{ugriz}$ for cluster members as follows:
\begin{equation*}
M_{ugriz} = m_{ugriz} - [\text{k-correction}] - [\text{distance modulus}].
\end{equation*}
We use the software package \verb+k-correct+ version 3.1 to perform 
this calculation \citep{Blanton:2007}. To calculate the distance 
modulus, we use a spectroscopic redshift when it is available, 
otherwise we use the cluster redshift, $z_\lambda$. Absolute 
magnitudes derived with this equation are used in Figures 
\ref{fig:sdss-lrg-cuts}, \ref{fig:sdss-lowz-cut} and 
\ref{fig:central-imag}.

Figure \ref{fig:sdss-lrg-cuts} illustrates how Equations
\eqref{eq:lrg-eq1} and \eqref{eq:lrg-eq2} operate. In this figure, we
show how cluster members and central galaxies from our sample populate
the $r_\text{Petro}$ versus $c_\|$ plane. There are additional cuts
that define LRGs (Equations \ref{eq:lrg-eq3}, \ref{eq:lrg-eq4},
\ref{eq:lrg-eq5}) but in practice these additional cuts only play a
minor role (see Table \ref{tb:LRG-cut}). The left panel in Figure 
\ref{fig:sdss-lrg-cuts} shows
cluster members color-coded by absolute magnitude and illustrates that
Equation \eqref{eq:lrg-eq1} (diagonal line) roughly selects a
population of LRGs brighter than an absolute magnitude limit of
$M_i\sim -23.2$. The right hand side of Figure \ref{fig:sdss-lrg-cuts}
shows redMaPPer centrals (the most likely central galaxies according
to redMaPPer). Roughly 25\% of redMaPPer centrals do not satisfy
Equations (\ref{eq:lrg-eq1}) - (\ref{eq:lrg-eq5}), mainly because they are
fainter than $M_i\sim -23.2$.

In order to obtain a volume-limited LRG sample, clustering studies
often apply an additional cut based on the $g$-band absolute
magnitude, $M_g$ \citep{Reid:2009,Zehavi:2005,Kazin:2010}. After this
$M_g$ cut, the LRG number density\footnote{Our values assume
  $h=0.7$. In $h$ inverse units, the number density of classical LRGs
  is $n_\text{LRG} = 1.0 \times 10^{-4} \ (h^{-1} {\rm Mpc})^{-3}$} is
$n_\text{LRG} = 3.4 \times 10^{-5} \ {\rm Mpc}^{-3}$. To make our
sample consistent with \citet{Reid:2009}, we apply a $M_g$ cut of
$-23.2<M_g<-21.2$ to the ``DR8PhotLRG'' sample to construct the
``DR8PhotLRG-Mgcut'' sample.  For consistency, we follow the same
method to compute $M_g$\footnote{When quoting values for $M_i$ we assume $h=0.7$. However, our $M_g$ values assume $h$ inverse units for consistency with \citet{Kazin:2010} and  \citet{Reid:2009}} as in
\citet{Kazin:2010}\footnote{http://cosmo.nyu.edu/$\sim$eak306/SDSS-LRG.html}:
\begin{equation}
\begin{split}
   \label{eq:abs-mg}
   M_g = r_\text{Petro} - &[\text{distance modulus}] - \Delta g \\
         &+ (g-r) - z_\text{calibration}.
\end{split}
\end{equation}

Since the observed $r$-band is close to the rest-frame $g$-band, $r_\text{Petro}$ can be used to predict $M_g$. $\Delta g$ and $(g-r)$ are color and k-corrections from Table 1 in \citet{Eisenstein:2001}, and $z_\text{calibration}=0.2$ accounts for evolution in the rest-frame colors from $z\sim0.3$ to $z=0$. We have checked that $M_g$ derived in this fashion and derived from \citet{Kazin:2010} yield similar results, despite the fact that \citet{Kazin:2010} used DR7 photometry and we use DR8. We find the difference in $M_g$ is small (RMS $\delta M_g = 0.07$), consistent with the scatter in the photometry and that the differences in $M_g$ do not affect the number of LRGs. Therefore, our results can be compared to results that directly use the \citet{Kazin:2010} LRG selection.

\subsection{BOSS LOWZ Photometric Cluster Member Selection}
\label{sec:lowz-selection}

In this section we explain how we construct our sample of BOSS ``LOWZ'' 
photometric cluster members (the ``PhotLOWZ'' sample). The details of 
the LOWZ selection can be found in \citet{Dawson:2013}. The main 
difference between classical LRGs and LOWZs is that the BOSS selection extends 
about 0.4 magnitudes deeper than the classical LRG selection. As a result, the 
LOWZ sample has a higher number density\footnote{Our values assume 
$h=0.7$. In $h$ inverse units, the LOWZ number density is $n_\text{LOWZ} 
= 3.0 \times 10^{-4} \ (h^{-1} {\rm Mpc})^{-3}$} than classical LRGs:  $n_\text{LOWZ} = 1.0
\times 10^{-4} \ {\rm Mpc}^{-3}$. In addition, the LOWZ selection 
includes a bright magnitude cut at $r_\text{cmod}>16$ which did not
exist in the classical LRG selection. Finally, the LOWZ selection
uses composite model magnitudes for magnitude cuts instead of Petrosian 
magnitudes.

Again, to avoid issues with fiber collisions, we define a photometric
LOWZ sample. The ancillary colors $c_{\perp,\text{LOWZ}}$ and 
$c_{\|,\text{LOWZ}}$ are different than the classical LRG selection 
criteria:
\begin{align}
    c_{\perp,\text{LOWZ}} 
    &= (r_\text{mod}-i_\text{mod}) - (g_\text{mod}-r_\text{mod})/4 - 0.18, \\
    c_{\|,\text{LOWZ}} 
    &= 0.7 (g_\text{mod}-r_\text{mod}) + 1.2 [(r_\text{mod}-i_\text{mod}) - 0.18].
\end{align}
With these new definitions we select PhotLOWZ galaxies using these cuts:
\begin{align}
   \label{eq:lowz-eq2}
   r_\text{cmod} &< 13.5 + c_\| / 0.3, \\
   \label{eq:lowz-eq3}
   16 &< r_\text{cmod} < 19.6, \\
   \label{eq:lowz-eq1}
   | c_\perp | &< 0.2, \\
   \label{eq:lowz-eq4}
   r_\text{psf} - r_\text{cmod} &> 0.3.
\end{align}
As in the classical LRG selection, the $c_\perp$ cut
crudely selects galaxies in the redshift range $z \lesssim 0.4$,
Equations \eqref{eq:lowz-eq2} and (\ref{eq:lowz-eq3}) give an absolute 
magnitude and an apparent magnitude cut, respectively,
and Equation \eqref{eq:lowz-eq4} is the star-galaxy separator. As
shown in Table \ref{tb:LOWZ-cut}, Equation (\ref{eq:lowz-eq2}) is the 
most stringent cut in defining the PhotLOWZ cluster member sample.

\begin{table}
\centering
\caption{Percentage of cluster members and central galaxies that fail
  to pass the LOWZ selection cuts given by Equations
  \eqref{eq:lowz-eq2} - \eqref{eq:lowz-eq4}. In total, before any cuts, 
  there are 507874 member galaxies and 7730 central galaxies. Among all 
  the cluster members, 479,040 (94.3\%) do not satisfy Equation 
  \eqref{eq:lowz-eq2} and 316,028 (62.2\%) do not satisfy Equation 
  \eqref{eq:lowz-eq3}. Applying all of these cuts,  24,761 cluster members (4.88\%) 
  and 7,001 central galaxies (90.6\%) are classified as LOWZs.}

\begin{tabular}{lcccc} \\ \hline
{\bf LOWZ cut}   &  Eq. \eqref{eq:lowz-eq2}  &  Eq. \eqref{eq:lowz-eq3}  
   &  Eq. \eqref{eq:lowz-eq1}  &  Eq. \eqref{eq:lowz-eq4}  \\ \hline
\shortstack{All cluster \\ members}
   &  94.3  &  62.2  &  5.37  &  4.74  \\ \hline
\shortstack{Most likely \\ central galaxies}
   &  7.84  &  1.54  &  0.285  &  0.0259  \\ \hline
\end{tabular}

\label{tb:LOWZ-cut}
\end{table}

\begin{figure*}
\centering
\includegraphics[width=16cm,clip]{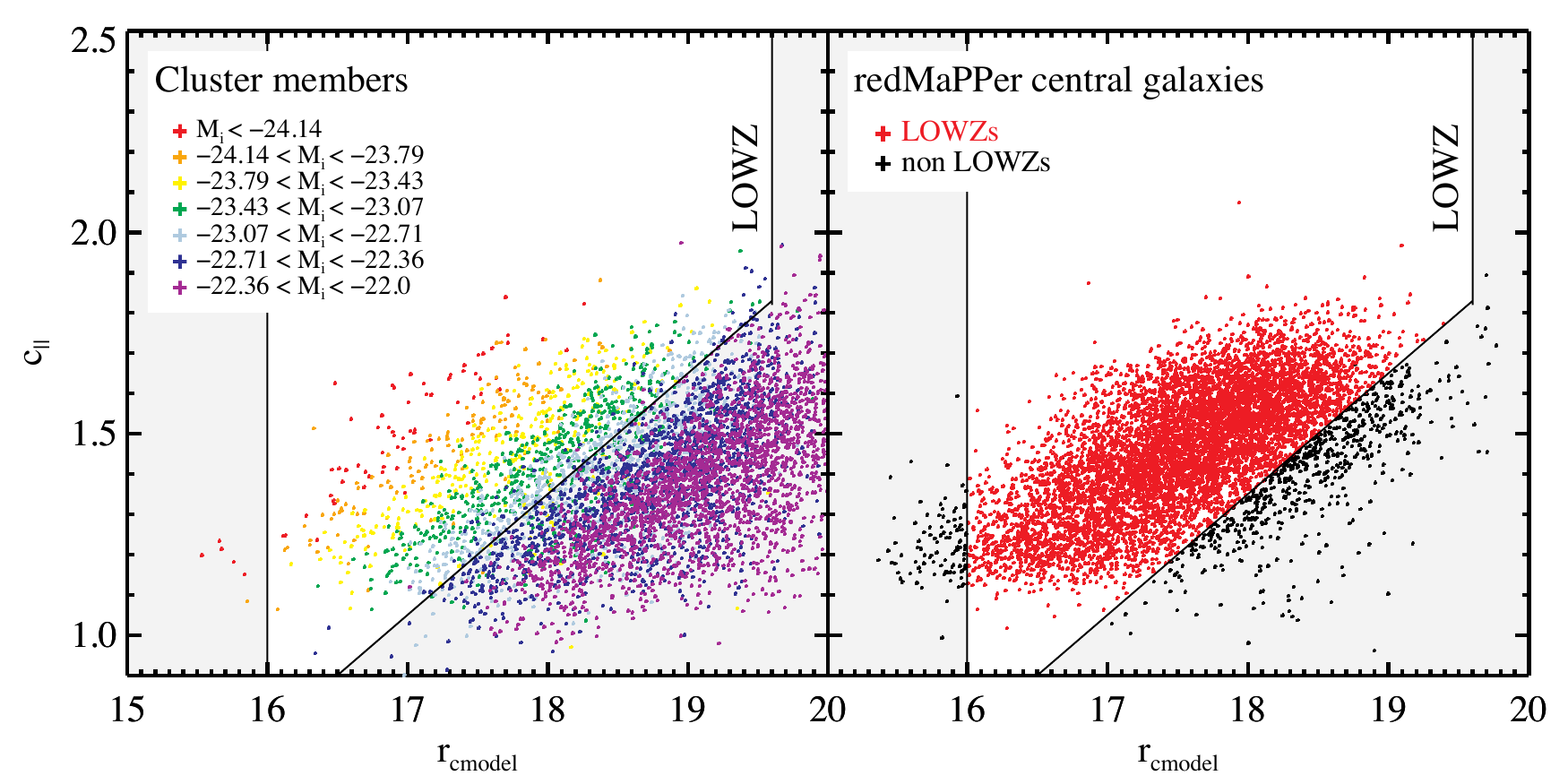}

\caption{Same as Figure \ref{fig:sdss-lrg-cuts} but for the BOSS
  LOWZ sample. Compared to classical LRGs, LOWZ  galaxies  extends
  to a fainter limit of $M_i \sim -22.8$. The right panel shows that
  the LOWZ selection includes a bright cut at $r_\text{cmod}>16$, which
  excludes 1.3\% of the central galaxies in our sample.}

\label{fig:sdss-lowz-cut}
\end{figure*}

In Figure \ref{fig:sdss-lowz-cut} we illustrate how the LOWZ selection
operates. As in Figure \ref{fig:sdss-lrg-cuts}, cluster members
color-coded by absolute $i$-band magnitude are shown on the left and
central galaxies are shown on the right. Absolute magnitudes are
calculated in the same way as in Section \ref{sec:classical-LRGs}. As
can be seen by comparing Figures \ref{fig:sdss-lrg-cuts} and
\ref{fig:sdss-lowz-cut}, the LOWZ selection reaches $\sim0.3$
magnitudes fainter in the $i$-band than the classical LRG
selection. At the bright end, 1.3\% (one hundred galaxies) of the most
likely redMaPPer central galaxies are rejected by the
$r_\text{cmod}>16$ cut.

\subsection{Selecting LRGs in clusters}
\label{sec:selections}

For most of our calculations, we consider all LRGs in a cluster with 
nonzero membership probability, which we will call the ``{\it 
probabilistic-sample}''. When using this sample, our calculations
will account for membership probabilities by summing over 
$p_\text{mem}$ values.

However, in some cases it is useful to have a deterministic sample of
LRGs in clusters (e.g., for plotting). In such cases, we select the
$\lambda$ most likely cluster members for each cluster ($\lambda$ is
rounded to the nearest integer). We will call this the ``{\it
  fixed-sample}.'' We will clarify in the text when each of these
samples is considered.  When using the fixed sample, the galaxy with
the highest value of $P_\text{cen}$ is called the redMaPPer central.

\section{Results}
\label{sec:results}

In this section we investigate two questions: (i) How often are central 
galaxies LRGs? and (ii) How often are central galaxies also the 
brightest LRGs in each cluster? For most of our calculations, we treat 
cluster centers and member galaxies probabilistically and use the 
centering probability, $P_\text{cen}$, and the membership probability, $p_\text{mem}$.

\subsection{Cluster Centroids}

We consider four different cluster centroids:

\begin{enumerate}
\item {\bf RMCG}: the redMaPPer central galaxy. For each cluster, 
this is the galaxy with the highest value of $P_\text{cen}$. This 
is the galaxy that redMaPPer selects as the most likely center 
for each cluster.
\vspace{0.5\baselineskip}

\item {\bf BMEM}: the brightest red-sequence cluster member identified
  by redMaPPer, where brightness is given by $i_\text{cmod}$. In most
  cases, this is defined with the {\it probabilistic-sample} and
  treated probabilistically according to the $p_\text{mem}$
  values. In some cases, BMEM will be defined using the {\it
    fixed-sample} (we clarify in each case which definition is used).
  \vspace{0.5\baselineskip}

\item {\bf BLRG}: the brightest DR8PhotLRG in each cluster (if the 
cluster has at least one DR8PhotLRG). This is defined in a similar fashion as 
BMEM.
\vspace{0.5\baselineskip}

\item {\bf BLOWZ}: the brightest PhotLOWZ in each cluster (if the 
cluster has at least 1 PhotLOWZ). This is defined in a  similar fashion as
BMEM.
\end{enumerate}

Note that RMCG is defined as the galaxy with the highest value of
$P_\text{cen}$, but that the actual centering probability of RMCG 
galaxies may be less than 1. In our sample, RMCGs have
$P_\text{cen}$ values between 0.27 and $\sim 1.00$ with a
mean value of $\langle P_\text{cen} \rangle=0.87$. Hence, RMCGs may
not always be the true cluster center. Indeed, redMaPPer predicts 
that 13\% of RMCGs will not be a true cluster center.

\begin{table}
\centering
\caption{Percent of time when two of samples (i) - 
(iv) overlap. For example, in clusters that have at least one LRG,
84.1\% of BLRGs  are also RMCGs. In clusters that have at least one
LRG/LOWZ, 94.7\% of BLOWZ are also BLRGs. Here, BLRG, BLOWZ and BMEM 
are defined using the ``{\it fixed-sample}''. }

\begin{tabular}{lcccc} \\ \hline
   &  RMCG  &  BLRG  &  BLOWZ  &  BMEM  \\ \hline
RMCG  &  100  &  84.1  &  80.1  &  80.5  \\ 
BLRG  &  -  &  100  &  94.7  &  95.5  \\
BLOWZ  &  -  &  -  &  100  &  96.3  \\
BMEM  &  -  &  -  &  -  &  100  \\ \hline
\end{tabular}

\label{tb:centroids}
\end{table}

We show a simple comparison between these different 
classifications in Table \ref{tb:centroids}. We use the ``{\it 
fixed-sample}'' (see Section \ref{sec:selections} for the sample 
definition) to define BLRGs, BLOWZs and BMEMs. 

\subsection{Central Occupation Functions of LRGs and LOWZ Galaxies}
\label{sect_hod}

We now compute the probability that a cluster of richness $\lambda$
hosts a central galaxy that is also an LRG or a LOWZ galaxy. When
expressed as a function of halo mass, this is known as the LRG/LOWZ
central occupation function, $N_\text{cen} (M_{200b})$. To begin
with, we compute $N_\text{cen}$ as a function of cluster richness. 
To measure $N_\text{cen} (\lambda)$, we use the ``{\it 
probabilistic-sample}'' where the sum of $P_\text{cen}$ for each
cluster is equal to 1.

We use 10 bins in cluster richness from $\lambda=20$ to $\lambda=150$. 
In each richness bin, we measure the mean value of $P_\text{cen}$ for
all LRG (or LOWZ) members. For any given cluster, the total value of
$P_\text{cen}$ summed over LRG members represents the probability that the
central galaxy of that cluster is an LRG. In the $j$-th bin,
$N_\text{cen} (\lambda)$ is:
\begin{equation}
   \langle N_\text{cen} (\lambda) \rangle_j
   = \frac{1}{n_{\text{clus}, j} (\lambda)} \sum_i P_{\text{cen},i}^\text{LRG}
\end{equation}
where $n_{\text{clus},j} (\lambda)$ is the number of clusters in the $j$-th
richness bin, and the subscript $i$ denotes all LRG cluster members in
this richness bin.

\begin{figure*}
\centering
\includegraphics[width=16cm,clip]{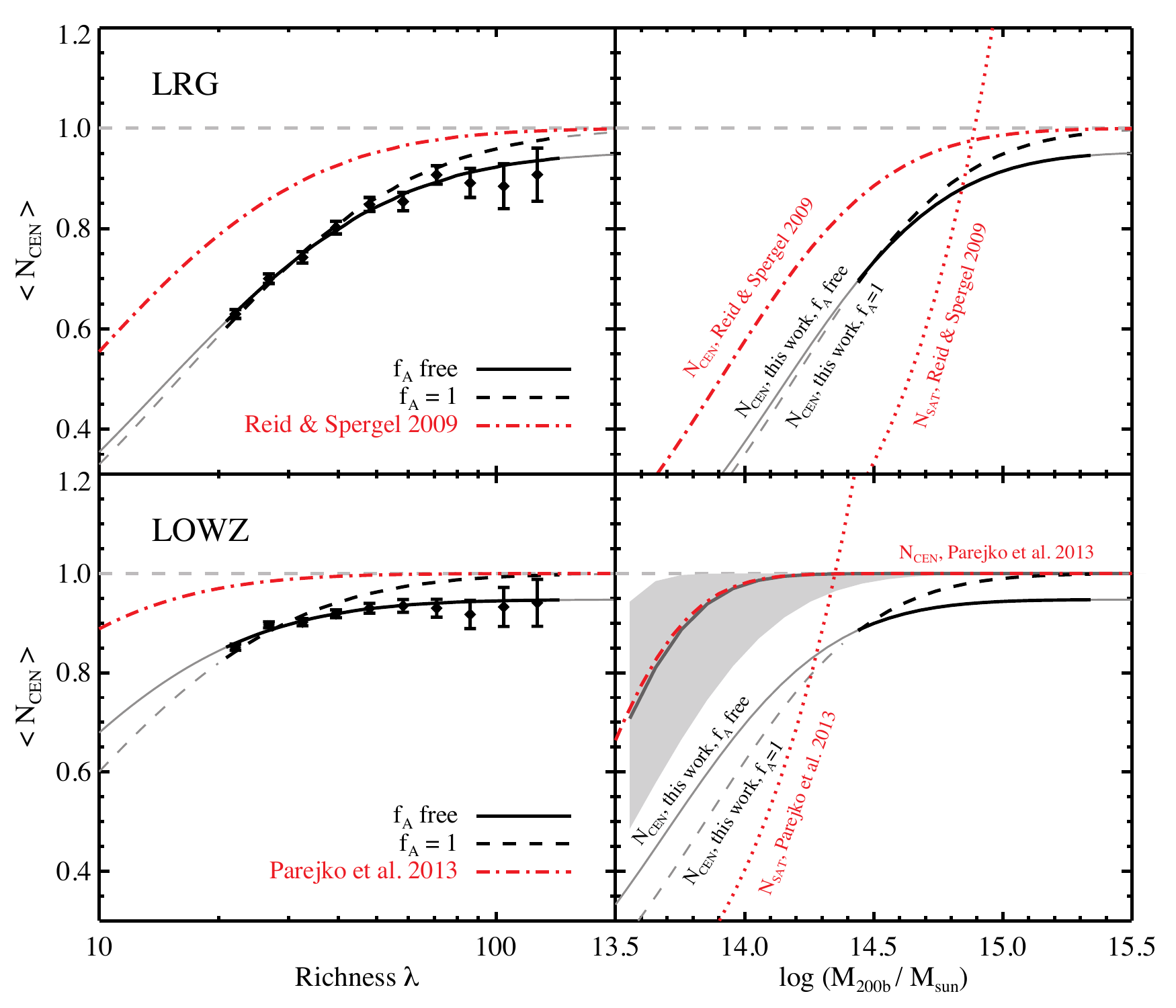}

\caption{{\bf Left panels}: $N_\text{cen} (\lambda)$ for LRGs (upper
  panel) and for LOWZ galaxies (lower panel). Solid black lines show
  our best fit to $N_\text{cen} (\lambda)$ when $f_A$ is left as a
  free parameter, and the black dashed lines show the best fit when
  $f_A=1$.  Red dash-dotted lines are the results from
  \citet{Reid:2009} and \citet{Parejko:2013} that have been converted
  to $N_\text{cen} (\lambda)$ using Equation
  \eqref{eq:ncen-lambda}. {\bf Right panels}: $N_\text{cen}
  (M_{200b})$ inferred by fitting Equation \eqref{eq:ncen-lambda} to
  the measured $N_\text{cen} (\lambda)$. Solid black lines show our
  results when $f_A$ is a free parameter. The thick black portion of
  the curve corresponds to the region constrained by the cluster
  data. Black dashed lines show our results when $f_A=1$. Red
  dash-dotted lines correspond to $\langle N_\text{cen} \rangle$ and
  red dotted lines correspond to $\langle N_\text{sat} \rangle$ from
  \citet{Reid:2009} (top) and \citet{Parejko:2013} (bottom). The grey
  solid line in the lower right hand panel shows $N_\text{cen}
  (M_{200b})$ we obtain by re-fitting the data points from
  \citet{Parejko:2013}. The 68 percent confidence region for this fit
  is shown using grey shaded regions. For both LRGs and LOWZs, we find
  a significant difference between the normalization of $N_\text{cen}$
  inferred from direct counts in clusters and that from HOD modeling of
  the two-point correlation function (compare black lines to red
  dash-dot lines).}
\label{fig:hod}
\end{figure*}

The results for $N_\text{cen} (\lambda)$ are shown in the two left panels 
of Figure \ref{fig:hod}. Errors are calculated via bootstrap. For 
consistency with \citet{Reid:2009}, we apply a $g$-band absolute 
magnitude cut to our LRG sample (we use the ``DR8PhotLRG-Mgcut'' sample, 
see Section \ref{sec:classical-LRGs}). For both LRGs and LOWZ galaxies, 
we find that $N_\text{cen} (\lambda)$ does not converge to 1 for large 
values of $\lambda$. Instead, $N_\text{cen} (\lambda)$ flattens and 
converges to $\sim 0.9$ at $\lambda=150$.

\subsubsection{Non-LRG redMaPPer centrals}

\begin{figure*}
\centering
\includegraphics[width=16cm,clip]{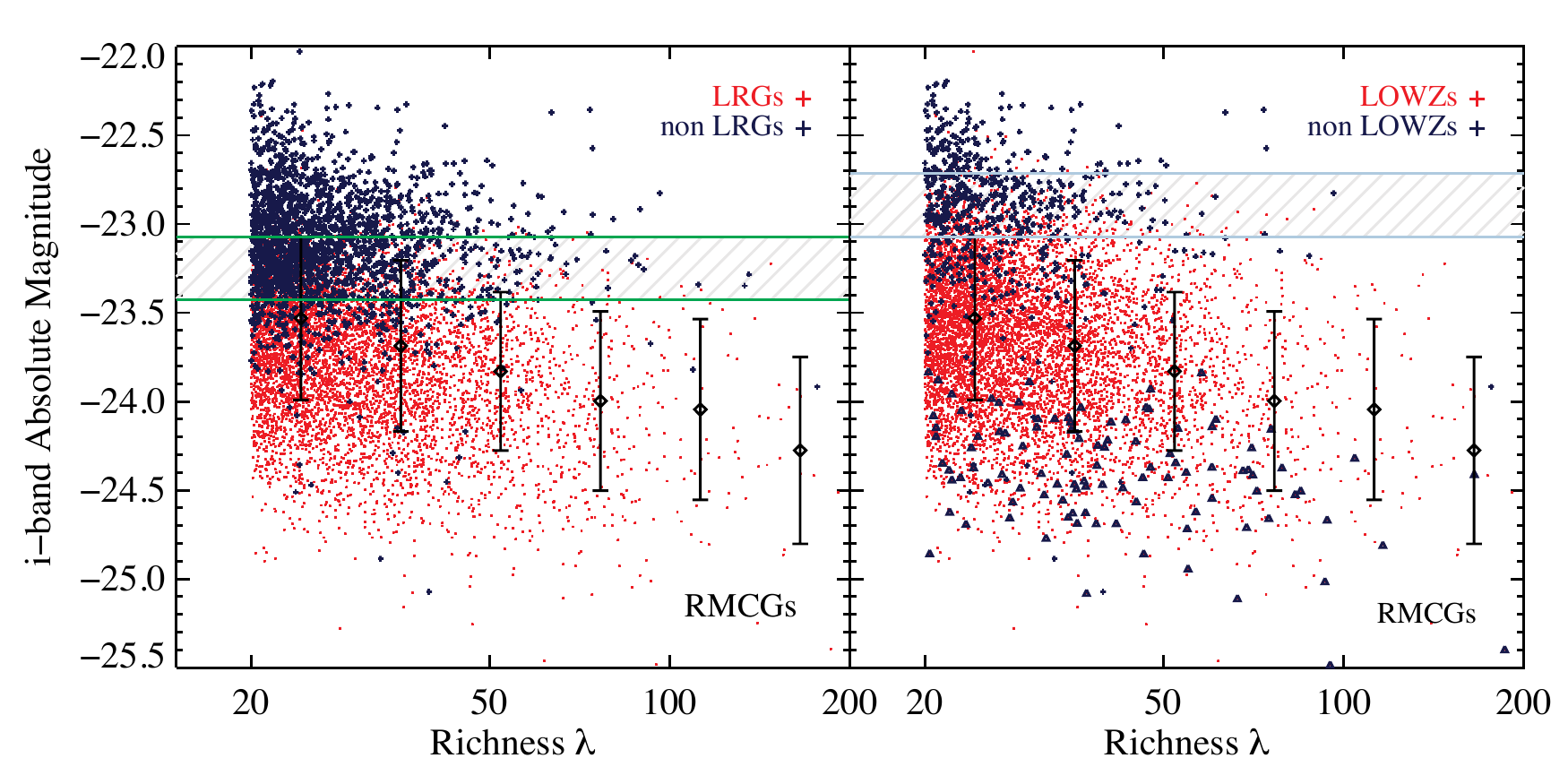}

\caption{{\bf Left panel}: $i$-band absolute magnitude of central
  galaxies as a function of cluster richness. Red points represent
  RMCGs that are also DR8PhotLRGs. Small dark blue points show RMCGs
  that fail to pass our photometric LRG selection. Open black diamonds
  with error bars show the mean magnitude in each richness bin with
  errors that represent the standard deviation of the distribution in
  each bin. The green shaded region is the approximate $i$-band
  magnitude limit that defines the LRG selection ($-23.43 \lesssim M_i
  \lesssim 23.07$, see Figure \ref{fig:sdss-lrg-cuts}). At
  $\lambda<40$, many RMCGs are simply too faint to pass the LRG cut,
  and there are excluded faint RMCGs at larger richness as well. {\bf
    Right panel}: similar plot as in the left panel using the LOWZ
  selection criteria. Red points represent RMCGs that are also
  PhotLOWZs, and black points represent RMCGs that are not
  PhotLOWZs. Filled blue triangles are RMCGs that are not PhotLOWZs because
  they are brighter than $r_\text{cmod} = 16$.  The light-blue
  horizontal band shows the absolute $i$-band magnitude cut that
  roughly defines the LOWZ selection ($-23.07 \lesssim M_i \lesssim
  -22.71$).}

\label{fig:central-imag}
\end{figure*}

The fact that $N_\text{cen} (\lambda)$ does not converge to 1 even for
large values of $\lambda$ can be explained by Figure
\ref{fig:central-imag}. 
This figure shows the $i$-band absolute magnitude distribution of the 
most likely RMCGs as a function of cluster richness. As expected, the 
absolute magnitudes are correlated with richness: brighter galaxies 
tend to be in richer clusters. However, there is significant 
scatter in this relation. The RMS dispersion of $M_i$ as a function of 
$\lambda$ is $\sigma_{\rm Mi} \sim 0.50$ for all RMCGs and 
$\sigma_{\rm Mi} \sim 0.46$ for RMCGs with $P_\text{cen} > 0.99$. The 
dispersion does not depend strongly on $\lambda$. These values are 
similar to those reported by \citet{Hansen:2005} for the MaxBCG 
cluster finder ($\sigma_{\log_{10} (Li)}=0.17$, or 
$\sigma_{\rm Mi}=0.425$). In Section \ref{sec:lrg}, we show that the 
LRG cuts roughly select galaxies above a fixed absolute magnitude 
threshold. This is shown by green horizontal lines in Figure 
\ref{fig:central-imag}. Because of the trend with $M_i$ and the
significant scatter in $M_i$, a sizable fraction (28\%) of RMCGs in
clusters with $\lambda < 40$ are simply too faint to pass the LRG
cuts. The scatter in $M_i$ at fixed halo mass also means there is a
non-negligible fraction of faint central galaxies in rich, massive halos
that fail to pass the LRG cuts.

The right hand side of Figure \ref{fig:central-imag} paints a similar
picture for LOWZ galaxies. The main difference is that the LOWZ
selection extends to fainter absolute magnitudes so a larger number of
RMCGs are also PhotLOWZs. There is however, one additional difference
due to the bright $r$-band cut applied for the LOWZ selection (see
Equation \eqref{eq:lowz-eq3} in Section
\ref{sec:lowz-selection}). This cut excludes about 102 bright central
galaxies (these have a mean $r$-band magnitude of $\langle
r_\mathrm{cmod} \rangle = 15.8$). For some studies, it may be
desirable to add these bright galaxies back in so as to remove the
effects of the LOWZ bright cut. Indeed, many of these very bright
galaxies are already targeted as main sample galaxies or as classical
LRGs\footnote{This will not be true for parts of the southern galactic
  cap that were extended by the BOSS survey.}. Among the 102 bright RMCGs
in our sample that do not pass the LOWZ cut, 73 have spectroscopic
redshifts in DR10. These missing galaxies contribute to the low
$N_{\text{cen}} (\lambda)$ at high halo mass shown in Figure
\ref{fig:hod}.

\subsubsection{Converting $N_\text{cen} (\lambda)$ to $N_\text{cen} (M_{200b})$}

We now convert $N_\text{cen} (\lambda)$ to $N_\text{cen} (M_{200b})$
to compare with results derived from HOD modeling of counts-in-cylinders
and the two point correlation function
\citep{Reid:2009,Parejko:2013}. In order to convert $\lambda$ to
halo mass, we adopt the richness-halo mass relation from
\citet{Rykoff:2012}. In Section \ref{sec:discussion}, we perform tests
to evaluate the impact of the uncertainty in this relation on our
results. In \citet{Rykoff:2012}, the probability that a cluster of
richness $\lambda$ has mass $M_{200b}$, $P(M_{200b}|\lambda)$,
is assumed to follow a log-normal distribution with a scatter of
$\sigma_{M|\lambda} = 0.25$. The mean of this log-normal relation is
given by:
\begin{equation}
   \label{eq:p-m200-lambda}
   \ln \left( \frac{M_{200b}}{h_{70}^{-1} 10^{14} M_\odot} \right)
      = 1.72 + 1.08 \ln \left( \frac{\lambda}{60} \right).
\end{equation}
Using this conversion, our lower richness limit of $\lambda =20$ 
corresponds to a halo mass of $M_{200b} = 2.44 \times 10^{14} 
M_\odot$ and a richness of $\lambda=100$ corresponds to a halo mass of 
$M_{200b} = 1.39 \times 10^{15} M_\odot$.

\citet{Rykoff:2012} specify $P(M_{200b}|\lambda)$ but 
$P(\lambda|M_{200b})$ is more useful for our calculations. We derive 
$P(\lambda|M_{200b})$ following the method described in the appendix 
of \citet{Leauthaud:2010} which assumes that $P(\lambda|M_{200b})$ 
is also a log-normal of the form:
\begin{equation}
   \label{eq:p-lambda-mh}
   P (\lambda | M_{200b})
   = \frac{1}{\sqrt{2\pi} \sigma_{\lambda|M}}
     \exp \left[ -\frac{(\ln \lambda - \ln \lambda_\text{mean} (M_{200b}))^2}
     {2 \sigma_{\lambda|M}^2} \right]
\end{equation}
For this calculation we adopt the halo mass function from
\citet{Tinker:2008}. From Equation \eqref{eq:p-lambda-mh} and the
halo mass function, we find $\sigma_{\lambda|M}=0.231$ and
\begin{equation}
   \ln \lambda_\text{mean} (M_{200b})
   = 0.925 \ln \left( \frac{M_{200b}}{M_\odot} \right) - 28.26.
\end{equation}
With this relation in hand, we now convert $N_\text{cen} (\lambda)$
to $N_\text{cen} (M_{200b})$. For this purpose, we assume that
$N_\text{cen} (M_{200b})$ follows the functional form:
\begin{equation}
   \label{eq:ncen}
   \langle N_\text{cen} (M_{200b}) \rangle
      = \frac{f_A}{2} \left[ 1 + \mbox{erf} \left( 
        \frac{\log_{10}( M_\text{200b}/M_\text{min})}{\sigma_{\log M}}
        \right) \right].
\end{equation}
This functional form for $N_\text{cen}$ is similar to that in
\citet{Reid:2009} but with an extra free parameter, $f_A$, which allows
$N_\text{cen}$ to converge to values less than unity at large halo
masses.

Given Equation \eqref{eq:p-lambda-mh} and assuming the
\citet{Tinker:2008} halo mass function $\frac{dn}{dm}$ at $z=0.2$, 
$N_\text{cen} (\lambda)$ is computed as :
\begin{equation}
   \label{eq:ncen-lambda}
   N_\text{cen} (\lambda)
   = \frac{\int P(\lambda|M_{200b}) N_\text{cen} (M_{200b}) 
     \frac{dn}{dm} (M_{200b}) dM_{200b}}
     {\int P(\lambda|M_{200b}) \frac{dn}{dm} (M_{200b}) dM_{200b}}
\end{equation}
We fit our measurements of $N_\text{cen} (\lambda)$ using Equation \eqref{eq:ncen-lambda} with free parameters $f_A$ and $M_\text{min}$. The best fit parameters are given in Table \ref{tb:ncen-params1}. The parameter $\sigma_{\log M}$ is fixed to 0.7, which corresponds to the value found by \citet{Reid:2009}. The cluster sample used in this paper ($\lambda>20$) does not extend to low enough halo masses to fully constrain $\sigma_{\log M}$. However, the \redmapper~ cluster catalog does extend to $\lambda>5$, which could help constrain $\sigma_{\log M}$ in future work.

\begin{table*}
\centering
\caption{HOD fits to $N_\text{cen}(\lambda)$ for the DR8PhotLRG-Mgcut
  sample and the PhotLOWZ sample. We fit $N_{\rm cen}(\lambda)$ both
  fixing $f_A=1$ and allowing $f_A$ to be a free parameter. The last
  two columns in this table are derived and discussed in Section
  \ref{sec:discussion}.}

\begin{tabular}{l|ccccc} \\ \hline
&   $f_A$   &   $\log (M_\text{min}/M_\odot)$
   &   $\chi^2/{\rm dof}$   &   $n_\text{CEN}$ [${\rm Mpc}^{-3}$]
   &   Implied satellite fraction  \\ \hline
LRG   &   1.0 (fixed)   &   $14.19\pm0.01$
   &   $15.1/9=1.7$  &  $1.8\times10^{-5}$  &   47\%
   \\ 
LRG   &   $0.953 \pm 0.014$   &   $14.14\pm0.02$
   & $4.87/8=0.61$   &  $1.9\times10^{-5}$  &  44\%
   \\ \hline
LOWZ   &   1.0 (fixed)   &   $13.85\pm0.01$
   &   $53.8/9=5.9$  &  $3.5\times10^{-5}$  &  65\%
   \\ 
LOWZ   &   $0.947\pm0.007$   &   $13.69 \pm 0.03$
   &   $15.2/8=1.9$  &  $4.2\times10^{-5}$  &  58\%
   \\ \hline
\end{tabular}

\label{tb:ncen-params1}
\end{table*}

\subsubsection{Comparison with \citet{Reid:2009} and \citet{Parejko:2013}}

In Figure \ref{fig:hod}, we compare our $N_\text{cen}(M_{200b})$ with
the central occupation function derived by \citet{Reid:2009}, who used
a combination of clustering and the counts-in-cylinders technique 
to constrain the HOD of LRGs. The halo mass definition assumed in
\citet{Reid:2009} is the same as this work\footnote{The HOD parameters given in \citet{Reid:2009} assume 
$h=0.7$. We have checked that integrating the \citet{Reid:2009} HOD with 
$h=0.7$ yields the correct LRG number density.}. The \citet{Reid:2009} 
results use halos identified by a spherical overdensity algorithm and so 
will be comparable with the \redmapper{} cluster finder.

For LOWZ, we compare with $N_\text{cen}$ derived by \citet{Parejko:2013}. 
Because \citet{Parejko:2013} use friend-of-friend halos, their results 
are not directly comparable to ours. We take two measures to account for 
this difference. First, we convert the \citet{Parejko:2013} results to 
our halo mass definition\footnote{ The FOF halo masses roughly correspond to an
overdensity of 500 times the mean matter density \citep{More:2011}. We 
convert the FOF halo masses from \citet{Parejko:2013} to our halo mass convention
assuming that $M_{200b}/M_{500b}=1.32$.} following \citet{More:2011}. 
Second, we refit the measurements of abundance and clustering of 
LOWZ galaxies presented by \citet{Parejko:2013}. For this, we use the 
analytical framework for the halo occupation distribution developed in 
\citet{van-den-Bosch:2013}, which accounts for 
the radial dependence of the bias, halo exclusion, and redshift space 
distortion effects on the projected clustering of galaxies \citep[see also][]{More:2013,
Cacciato:2013}. In fitting 
the clustering data, we use the parameterization as 
\citet{Parejko:2013} \citep[see also][]{Miyatake:2013,More:2014}. We adopt 
a consistent halo mass definition of $M_{200b}$ while performing the 
fits so that the HOD modeling results can be directly compared with the 
results from our cluster catalog. The 68 and 95 percent confidence 
regions are shown using grey shaded regions in Figure \ref{fig:hod}. Our 
re-fits to the \citet{Parejko:2013} data points are in excellent 
agreement with \citet{Parejko:2013} HOD after converting to a single halo mass 
definition, suggesting that differences between friend-of-friend halos 
and spherical overdensity halos do not have a large impact on
$N_\text{cen}$.

For both the LRG sample and the LOWZ sample, we find significant 
differences in the amplitude of $N_\text{cen}$ derived from HOD modeling 
of LRG/LOWZ clustering and derived directly from the redMaPPer cluster 
catalog. Possible explanations for this discrepancy are presented in 
Section \ref{sec:discussion}.

\subsection{How often is the Brightest LRG not the Central Galaxy?}
\label{sec:rmcg-blrg}

Although many studies assume that the brightest galaxy in a
cluster is also the central galaxy \citep{van-den-Bosch:2004, 
Weinmann:2006, Budzynski:2012}, some recent studies suggest that is 
not always the case \citep{van-den-Bosch:2005aa, Coziol:2009aa, 
Sanderson:2009aa, Hikage:2013}. Because
redMaPPer uses color and position in addition to luminosity to
determine which galaxy is the central galaxy, we can test this
assumption by calculating the probability that the brightest
LRG/LOWZ galaxy in a cluster is not the central galaxy (denoted
$P_\text{BNC}$). To compute $P_\text{BNC}$, we follow the
probabilistic method outlined in Reddick et al., in prep. For each
cluster, we identify all LRGs brighter than the $i$-th candidate
central. $P_\text{BNC}$ for LRG is computed as:
\begin{equation}
   \label{eq:Pbnc}
   P_\text{BNC} = \sum_{i=1}^5 P_{\text{cen}, i}
                  \sum_j  P_{\text{mem},j}
                  \prod_k (1-P_{\text{mem},k}).
\end{equation}
The first sum runs over all five candidate centrals, not limited to
LRGs. The second sum over $j$ is for all LRGs that are brighter than
central candidate $i$. The final product over $k$ is for
LRG members brighter than galaxy $j$ to ensure that only the
BLRG is considered.

\begin{figure}
\centering
\includegraphics[width=8cm,clip]{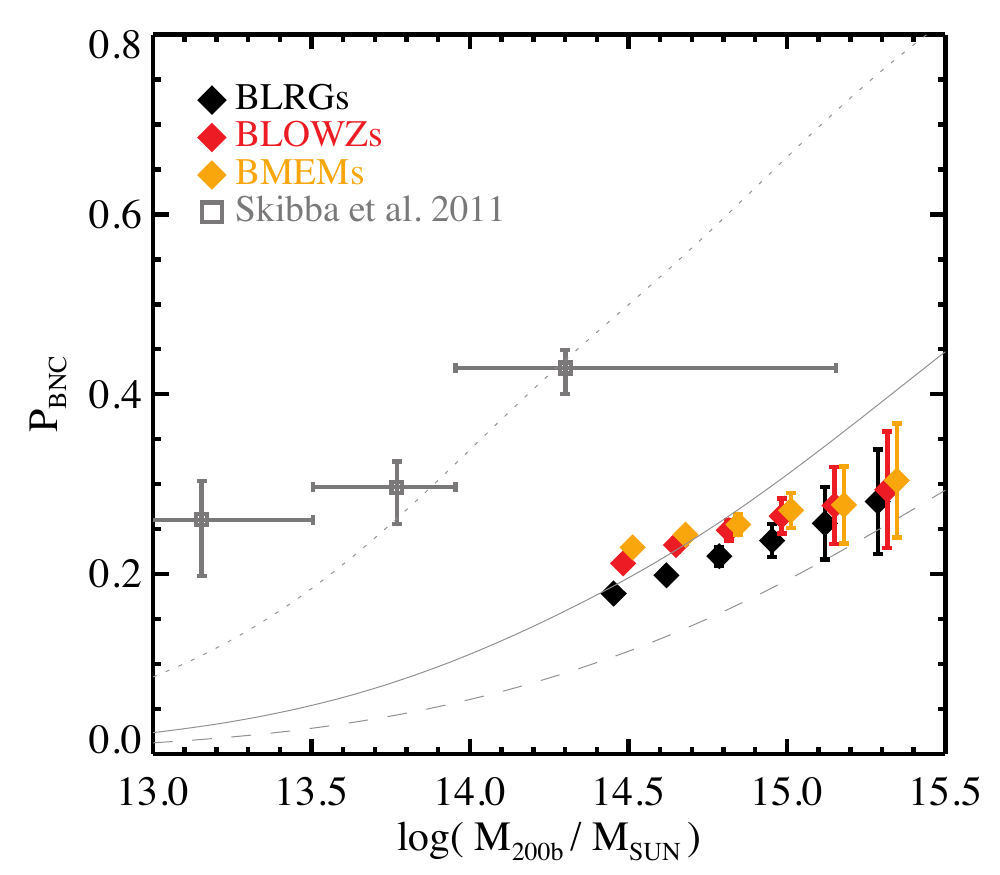}

\caption{Probability that there exists a redMaPPer cluster member (orange), an LRG
member (black), or a LOWZ member (red), that is brighter than the central galaxy
as a function of halo mass. Errors are derived via bootstrap. Grey
points show measurements from \citet{Skibba:2011}. Grey lines show
predictions from \citet{Skibba:2011} based on the conditional luminosity 
function (CLF) from Cacciato et al. 2009, and assuming three different 
values for the slope in the satellite CLF ($s= 1$ dotted curve, $s = 2$ 
solid curve, $s = 3$ dashed curve)}

\label{fig:blrg-pcen}
\end{figure}

\begin{figure}
\centering
\includegraphics[width=8cm,clip]{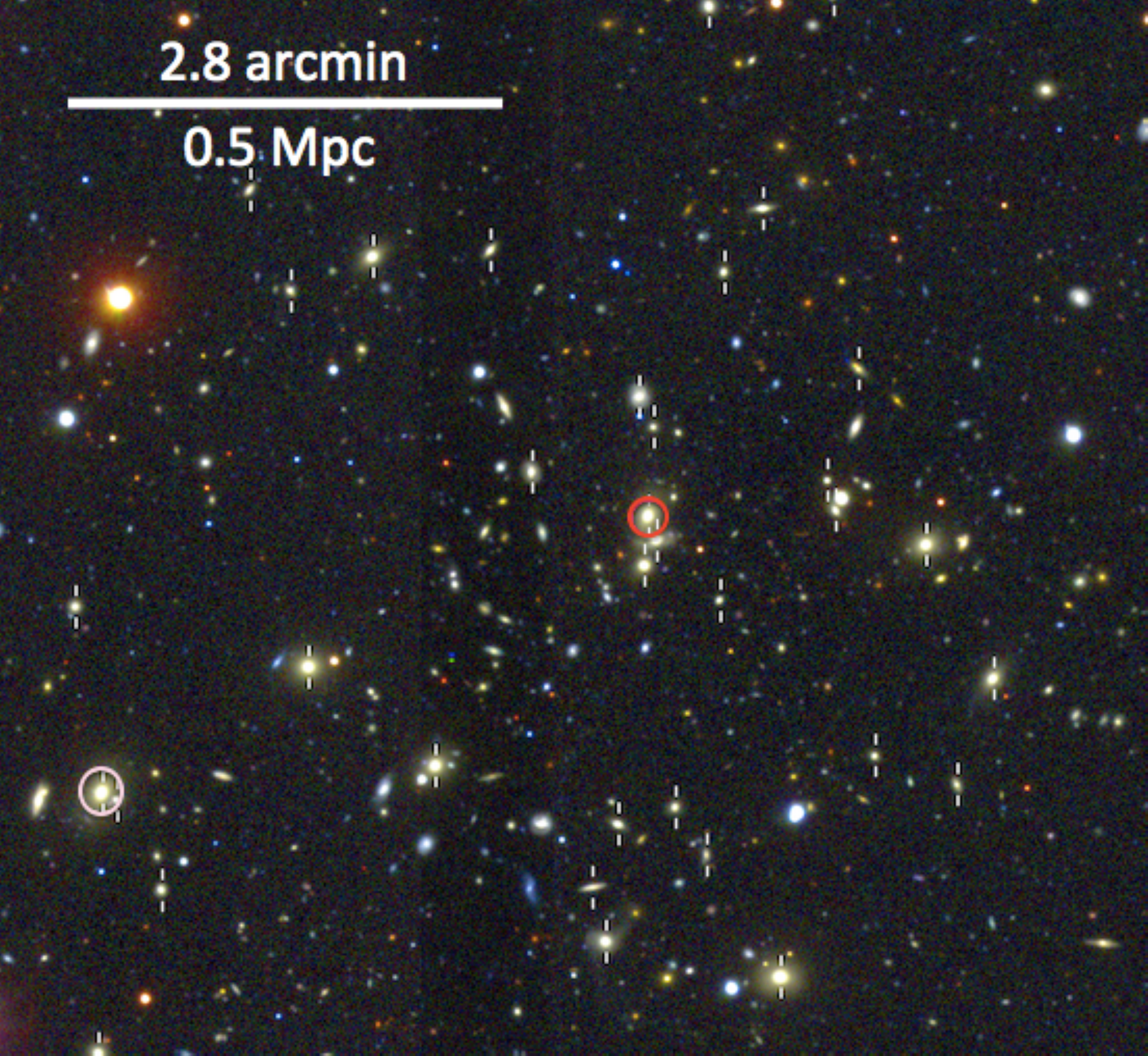}
\caption{SDSS image of a cluster in which the redMaPPer central galaxy
  is not the brightest member. Member galaxies are shown with white
  ticks. The richness of this cluster is $\lambda=54.8$. The galaxy
  circled in red in the middle is the most likely central, and the one circled in
  pink (lower left) is the brightest member. The centering probability
  of the central galaxy is 0.930. The brightest member of this cluster
  has $P_\text{cen} = 0.0$. The $i$-band apparent magnitudes of the
  central galaxy and the brightest member are $m_i=16.8$ and
  $m_i=16.3$, respectively.}
\label{fig:cluster-image}
\end{figure}

Figure \ref{fig:blrg-pcen} shows the mean probability that the BLRG or
BLOWZ galaxy in a cluster is not the central galaxy. We find that
$P_\text{BNC}$ varies between 0.2 and 0.3 and mildly increases with
halo mass. Figure \ref{fig:cluster-image} shows an example of a
cluster in which the central galaxy is not the brightest member. The
most likely central galaxy of this cluster has $P_\text{cen}=0.930$,
and the $P_\text{cen}$ value for the brightest galaxy (in the {\it
  fixed-sample}) is 0. The central is not an LRG but the brightest
member is an LRG. In this cluster, the BLRG is less likely
to be the central because it is offset from most other cluster members. 

We compare our results with \citet{Skibba:2011} in Figure \ref{fig:blrg-pcen}. \citet{Skibba:2011} compute $P_\text{BNC}$ by comparing the velocity and positional offsets of the brightest cluster galaxy from the other cluster members to the expected offsets computed using mock group catalogs. The \citet{Skibba:2011} results are derived from the \citet{Yang:2007aa} group catalog. The details of the cluster finding algorithm are accounted for in the \citet{Skibba:2011} analysis by running the \citet{Yang:2007aa} group finder on mock catalogs. Note that the \citet{Skibba:2011} results span a lower redshift range ($0.01 < z < 0.20$), and larger halo mass range ($12 < \log(M_\text{halo}/M_\odot) < 15$) than our results. We convert the \citet{Skibba:2011} results to our assumed value of $h=0.7$. We do not, however, account for the difference in halo mass definition, $M_{180b}$ in \citet{Skibba:2011} versus $M_{200b}$ here, as the correction is of order $3\%$.

Our values of $P_\text{BNC}$ are significantly lower than those found
by \citet{Skibba:2011}. At a halo mass of $\log (M_{180b}/M_\odot) =
14.29$, \citet{Skibba:2011} find $P_\text{BNC}=0.43$, whereas we find
$P_\text{BNC} = 0.23$ at $\log (M_{200b}/M_\odot)=14.5$. As
demonstrated in Reddick et al., in prep, part of the difference may be
explained by a tendency to over-subtract the sky around bright,
extended galaxies in SDSS data releases prior to DR8. While we rely on
DR8 data, \citet{Skibba:2011} use SDSS DR4 photometry. Improvements in
the sky subtraction in DR8 typically increased the measured
luminosities of central galaxies more than those of satellites
\citep{Bernardi:2013}, thus decreasing the probability that a
satellite is brighter than the central galaxy. However, the sky
subtraction is most relevant at redshifts below the range we study
here, and may only explain part of the discrepancy. We refer to Reddick
et al., in prep for further details on this question.

In conclusion, we find that within the redMaPPer cluster catalog, the
brightest LRG/LOWZ galaxy is also the central galaxy in only
$\sim$70\% of clusters.  This will impact redshift space distortions
(RSD) studies that construct ``halo catalogs'' under the assumption
that the brightest LRG is always the central galaxy
\citep{Hikage:2013aa}. This also suggests that naive cluster-finding
algorithms that assume the central galaxy is the brightest galaxy in a
cluster may have a $20 \sim 30\%$ mis-centering fraction
\citep{van-den-Bosch:2005aa, Coziol:2009aa, Sanderson:2009aa}.

\subsection{Offset Distributions between the Central Galaxy and the 
Brightest LRG}
\label{sec:offset-distribution}

We now investigate how the radial distributions of non-central BLRGs compare with the radial distributions of red cluster members. We compute the distribution of projected offsets between the brightest and central galaxies for clusters in which the BLRG is not the central galaxy. Specifically, we compute $D_\text{off} / R_\lambda$, where $D_\text{off}$ represents the 2D projected distance between the central galaxy and the BLRG/BLOWZ. We normalize $D_\text{off}$ by the cluster cutoff radius, $R_\lambda$, to account for size variations among clusters. $D_\text{off}$ is computed as follows. For each cluster, we consider all 5 central galaxy candidates. Let us consider the $i$-th central candidate with a centering probability of $P_{\text{cen}, i}$. We identify all LRG cluster members brighter than central galaxy $i$. The probability of observing an offset $D_{\text{off}, ij}$ between central galaxy $i$ and the brightest LRG $j$ in the cluster is:
\begin{equation}
   \label{eq:p-doff}
   P(D_{\text{off}, ij}) = P_{\text{cen}, i}  P_{\text{mem},j} \ 
       \prod_k (1-P_{\text{mem},k}),
\end{equation}
which is simply the $i,j$-th term in the sum making $P_\text{BNC}$
(see Equation \ref{eq:Pbnc}).

\begin{figure}
\centering
\includegraphics[width=8cm,clip]{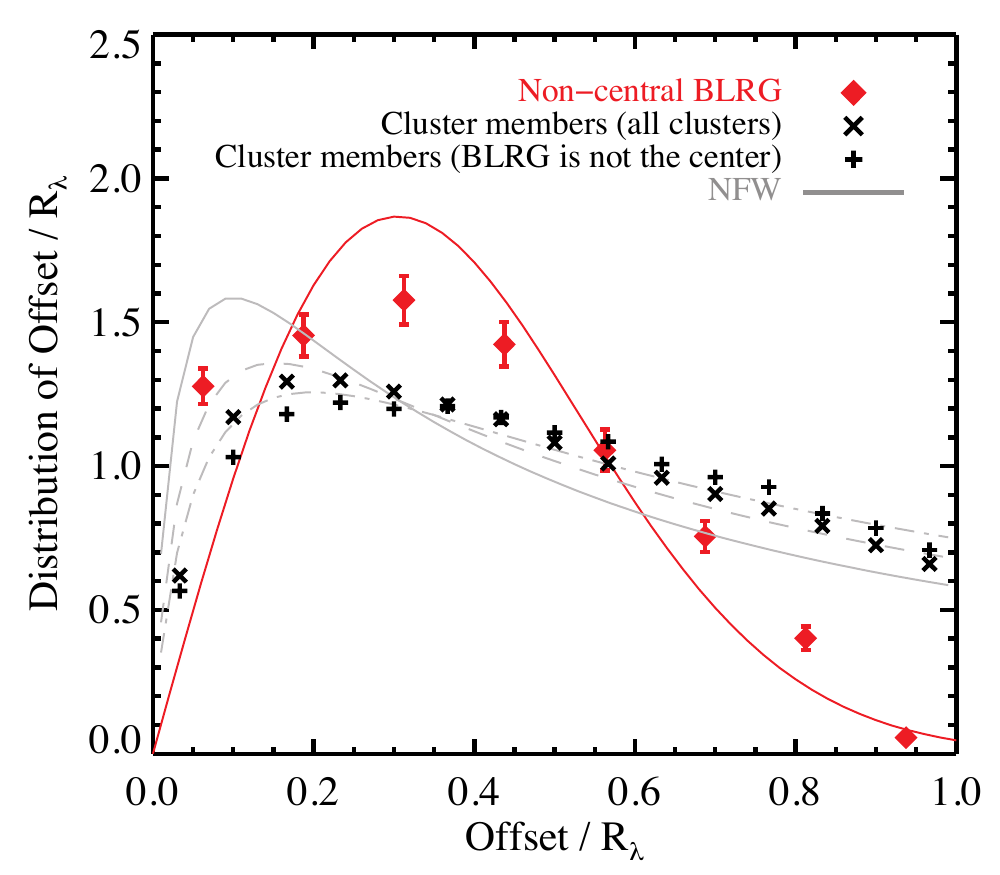}

\caption{Distribution of $D_\text{off}/R_\lambda$ for the brightest
  LRGs, when the RMCG and BLRG are different. Errors are calculated by
  bootstrap. The red solid curve shows the best fit assuming a Rayleigh distribution.  Black crosses represent the distribution of
  all \redmapper{} cluster members. Black plus signs represent the
  distribution of all \redmapper{} cluster members in clusters in
  which the BLRGs are not the central galaxies. The fact that the
  black crosses and black plus signs trace similar distributions
  suggests that clusters with non-central BLRGs are not dominated by
  projection effects and/or mergers. Grey curves represent the
  expected distribution of satellites distributed according to NFW
  profiles with different halo masses at redshift $z=0.25$. Grey
  solid, dashed, and dash-dotted curves are
  $\log(M_{200b})=14.0,14.5,15.0$, respectively (and the halo
  concentrations are $c=7.2,6.5,5.3$). The distribution of
  $D_\text{off}/R_\lambda$ for non central BLRGs is significantly
  different (truncated at the outskirts) compared to the distribution for all cluster
  members.}
\label{fig:distance-offset}
\end{figure}

 Figure \ref{fig:distance-offset} is constructed by finding all
 configurations in which LRG $j$ is brighter than central galaxy $i$
 and computing $D_{\text{off}, ij}$ and $P(D_{\text{off}, ij})$. We then
 bin the $D_\text{off}/R_\lambda$ values and sum the $P(D_{\text{off},ij})$ 
 values in each bin. We only show the distribution of offsets
 for BLRGs, because the distributions are similar for BLOWZ galaxies
 and BMEM galaxies. The mean offset between the RMCG and the brightest
 LRG (when these galaxies differ) is $0.4 R_\lambda$.

 For reference, we show the expected distribution of $D_\text{off} /
 R_\lambda$ assuming an NFW profile (grey lines). This is computed by
 first integrating the three dimensional density profile along the
 line-of-sight to obtain the projected surface mass density. The
 curves are then computed by multiplying the area of a thin annulus at
 each radius to the density at each radius. Each grey line shows a
 different halo mass and halo concentration. Figure
 \ref{fig:distance-offset} shows that the distribution of 2D projected
 halo-centric distances to BLRGs is significantly different (more
 truncated at the outskirts) than the distribution of radial distances
 to all cluster members. The differences between these distributions
 may indicate a larger impact of dynamical friction on more massive
 sub-halos as suggested by \citet{Wu:2013}. 
 Figure 7 also shows that the distribution of radial offsets for
 member galaxies in clusters is the same whether or not the BLRG is
 the central galaxy. This suggests that clusters with non-central
 BLRGs are not dominated by odd (e.g. merging) clusters.

Previous work has sometimes modeled the distribution of offsets using 
a Rayleigh distribution function instead of an NFW profile 
\citep{Johnston:2007aa}:
\begin{equation}
   P(d) = \frac{d}{\sigma^2} \exp \left[ -\frac{1}{2} \left( 
          \frac{d}{\sigma} \right)^2 \right]\ ,
\end{equation}
where $d=D_\text{off}/R_\lambda$. As shown in Figure \ref{fig:distance-offset}, a Rayleigh distribution function with $\sigma=0.39$ (our best fit value) does not provide a satisfactory description of this distribution, although it provides a better fit than an NFW distribution. 

\section{Discussion on the origin of the $N_\text{cen}$ discrepancy}
\label{sec:discussion}

The above results show that (i) the fraction of \redmapper{} clusters
with a central LRG ($N_\text{cen}(\lambda)$) does not converge to unity 
for rich clusters, (ii) the brightest LRG in a cluster is not the
central galaxy in $20 \sim 30\%$ of cases, and (iii) the offset distributions of 
non-central BLRGs is significantly different than that of all cluster 
members. Below, we put forth possible
explanations for why our measured $N_\text{cen}$ is lower than that
inferred from clustering. 
 
\subsection{Possible (but unlikely) explanations} 
\label{ssec:possible-explains}

One reason that our values for $N_{\text{cen}}$ are lower than those
found in previous studies might be that the \redmapper{} algorithm
does not correctly identify central galaxies. We have conducted a
visual inspection of 272 \redmapper{} clusters in our redshift
range. We created color images for these clusters using the Stripe 82
coadds \citep{Annis:2014aa} which are $\sim2$ mags deeper than the
SDSS single epoch imaging.  Five co-authors inspected these clusters
to select ``visual" central galaxies. Our visual inspection did not
reveal any obvious failure modes in the \redmapper{} central galaxy
selection and we found good agreement between the ``visual" central
galaxies and the \redmapper{} galaxy with the highest value of
$P_\text{cen}$ (95\% of the time).  From these tests alone we cannot
rule out the possibility that the differences that we find for
$N_\text{cen}$ are due to an incorrect central galaxy assignment by
\redmapper{}. However, our visual inspections did not reveal any
obvious issues. In Paper II we perform tests of the redMaPPer
centering probabilities using weak gravitational lensing and
cross-correlations. Early results from this work suggest that the
redMaPPer centering scheme outperforms other centering choices, such
as the bright cluster member.

Based on the above tests, we think that mis-centering in the \redmapper{} cluster catalog is unlikely to explain the $N_\text{cen}$ discrepancy. Another possibility is that part of the difference may be caused by uncertainties in the \citet{Rykoff:2012} mass-richness relation. Indeed, the \citet{Rykoff:2012} mass-richness relation was derived for an older version of the \redmapper{} cluster catalog. To determine how large of an impact this might have, we use an updated calibration from Reddick et al. in prep, which is based on the same \redmapper{} catalog as the one used here\footnote{Because Reddick et al. in prep assume a different cosmology than we do, we recalibrate their mass-richness relation for this test. This re-calibration yields the same cluster abundances as a function of $\lambda$ as Reddick et al. in prep.}. We re-derive our results using the updated mass-richness relation but as shown in Figure \ref{fig:ncen-test}, this only slightly alters our results. Finally, we also vary $\sigma_{M|\lambda}$ between $\sigma_{M|\lambda} = 0.22$ and $\sigma_{M|\lambda} = 0.28$ but this does not impact our conclusions. Based on these tests, we conclude that uncertainties in the \citet{Rykoff:2012} mass-richness relation are unlikely to explain the discrepancy.

Another possibility is that there is a systematic difference between
our photometric LRG selection and the actual spectroscopic samples.
However, we obtain similar values for $N_\text{cen} (\lambda)$ if we
select LRGs based on DR7 photometry, or based on the original LRG
targeting flag (\verb+TARGET_GALAXY_RED+, see Appendix A).  As discussed
in the Appendix A, differences between our photometric LRG selection and
\verb+TARGET_GALAXY_RED+ are consistent with scatter in the photometry
between reductions. Hence, differences between the LRG samples from
\citet{Reid:2009} and \citet{Parejko:2013} and our photometric LRG
selection are unlikely to be the source of the $N_{\text{cen}}$
discrepancy.

Another possibility may be related to issues in our
  conversion from $N_\text{cen}(\lambda)$ to $N_\text{cen}
  (M_{200b})$. For example, for consistency with \citet{Reid:2009}, we
  fixed the parameter $\sigma_{\log M}$ to 0.7. However, we have also
  tested fits with $\sigma_{\log M}=0.5$ and $\sigma_{\log
    M}=0.9$. The shape of $N_\text{cen} (M_{200b})$ is modfied as
  expected but the overall discrepancy persists. The conversion of
  $N_\text{cen}(\lambda)$ to $N_\text{cen} (M_{200b})$ also depends on
  the assumption that $P(M_{200b}|\lambda)$ follows a log-normal
  distribution. While this is a very common assumption, it is by no
  means a well tested one. It would be interested to test how this
  assumption would affect our results, but we defer this aspect to
  future work. Finally, this conversion may also fail if redMaPPer
  clusters do not represent a perfect recovery of dark matter
  halos. If the cluster catalog contains purity and/or completeness
  issues, then it will not be safe to assume that one can derive
  $N_\text{cen} (M_{200b})$ from
  $N_\text{cen}(\lambda)$. Completeness, however, does not appear to
  be an issue for the \redmapper{} cluster catalog, beyond the
  expected scatter in the richness-mass relation, which we have
  already accounted for in the density matching procedure. Purity
  however, is more of a concern.  Preliminary X-ray follow-up of a
  complete sample of \redmapper{} clusters suggests that the fraction
  of redmapper clusters that are projections is no larger than 7\%. A
  7\% change in density corresponds to a $\sim$2\% change in mass
  which corresponds to a $\sim$2\% change in the scaling relation
  \citep{Weinberg:2013} and should not dramatically alter our
  conclusions.

\subsection{Implication of our Results in Terms of Central and
  Satellite Number Densities} 

In order to compare our results to previous work based on clustering, we compute the number density of central and satellite galaxies implied by our measurements of $N_\text{cen} (\lambda)$. To compute the central number density, we integrate the product of the best-fit $N_\text{cen} (M_{200b})$ and $\frac{dn}{dm} (M_{200b})$ over the full halo mass range. For this calculation, we are forced to extrapolate our model for $N_\text{cen} (M_{200b})$ to lower halo masses that are not sampled by the redMaPPer galaxy clusters used here. There is a potentially large systematic uncertainty associated with this extrapolation. On the other hand, the {\it erf} model in Equation \eqref{eq:ncen} is the standard for HOD models. Therefore, any discrepancies between our work and clustering studies that can be traced back to the extrapolation of $N_{\text{cen}}(M_{200b})$, would correspond to a failure in the now standard HOD model. Thus, we simply caution that our conclusions are dependent on the validity of the {\it erf} extrapolation. Also note that the goal of this
exercise is simply to gain a sense of the differences between our
results and previous work based on clustering.

Our HOD implies a number density of central galaxies of
$1.9\times10^{-5} \ {\rm Mpc}^{-3}$ for LRGs and a number density of
$4.2\times10^{-5} \ {\rm Mpc}^{-3}$ for LOWZ galaxies (see Table
\ref{tb:ncen-params1}). Assuming a total number density of $n_\text{TOT} = 3.4 \times 10^{-5} \ {\rm Mpc}^{-3}$ for LRGs and $n_\text{TOT}=1.0\times10^{-4} \ {\rm Mpc}^{-3}$ for LOWZ galaxies, we can
infer the number density of satellite galaxies ($n_\text{SAT} = n_\text{TOT}-n_\text{CEN}$) and the implied satellite fraction (see Table
\ref{tb:ncen-params1}). For LRGs we find that our model implies a
satellite number density of $1.5 \times 10^{-5} \ {\rm Mpc}^{-3}$
which corresponds to an LRG satellite fraction of 44\%. Likewise, for
LOWZ, we find that our model implies a satellite number density of
$5.8 \times 10^{-5} \ {\rm Mpc}^{-3}$ which corresponds to a satellite
fraction of 58\%.  These satellite fractions are much higher those of
\citet{Reid:2009} and \citet{Parejko:2013}, which are 6.4\% and 12\%,
respectively.

The fact that our HOD fits to $N_\text{cen} (M_{200b})$ imply much
larger satellite fractions than previous works is not surprising
because our $N_\text{cen} (M_{200b})$ is lower than those of
\citet{Reid:2009} and \citet{Parejko:2013}. Hence, the satellite
fraction must increase in order to preserve the total LRG number
density. The satellite fractions derived in this section depend on
multiple assumptions and on extrapolating below our halo mass
limit. Nonetheless, the satellite fractions inferred from our fits are
uncomfortably high --- leading us to explore effects that might
invalidate our measurements of $N_\text{cen} (M_{200b})$. Since our
measurements of $N_\text{cen} (\lambda)$ are reliable (see
\S\ref{ssec:possible-explains}), we focus on effects that might
influence the conversion from $N_\text{cen} (\lambda)$ to
$N_\text{cen} (M_{200b})$.

\subsection{How would our Results be Affected by a Strong Correlation
  Between Richness and Central Luminosity?} 

Our results rely on converting $\lambda$ to $M_{200b}$ via Equation
\eqref{eq:p-m200-lambda}, which assumes that $\lambda$ only depends on
$M_{200b}$. Here, we consider the possibility that $\lambda$ also correlates with $M_I^{\rm CG}$, the absolute magnitude of the central galaxy. 

In this case, the
conversion in Equation \eqref{eq:p-m200-lambda} will depend on
$P(\lambda|M_{200b},M_I^{\rm CG})$ instead of just
$P(\lambda|M_{200b})$. To determine how this correlation might impact
our results, we evaluate $N_\text{cen} (M_{200b})$ for two mock samples in which halo mass and central galaxy luminosity are maximally correlated and anti-correlated, respectively, at fixed richness. We design these samples in the following way:

\begin{enumerate}
\item We build a mock catalog that contains halos drawn from the
  \citet{Tinker:2008} mass function. Each halo is assigned a richness
  based on Equation \eqref{eq:p-lambda-mh}.  \vspace{0.5\baselineskip}

\item We assign an $i$-band central galaxy absolute magnitude to mock
  halos. Mock $M_I^{\rm CG}$ values are randomly drawn from the real
  redMaPPer catalog after matching mock clusters and real clusters by
  richness. This procedure ensures that our mock catalog has the same
  relationship between $\lambda$ and $M_I^{\rm CG}$ as the
  \redmapper{} catalog.  \vspace{0.5\baselineskip}

\item We then bin the clusters by $\lambda$. Each narrow richness
  bin contains a range of halo masses and central galaxy
  magnitudes. In the initial catalog, there is no correlation between
  $M_I^{\rm CG}$ and $M_{200b}$. Within each bin, we will
  \emph{reassign} the $M_I^{\rm CG}$ values in order to introduce a
  correlation between halo mass and central galaxy magnitude.
\vspace{0.5\baselineskip}

\item In our first model (MOD1), we reassign all central galaxies such that
  heavier halos contain brighter centrals within each richness bin (i.e., $M_{200b}$ and $M_I^{\rm CG}$ are maximally correlated at fixed $\lambda$). In this
  model, clusters with lower values of $\lambda$ have brighter
  central galaxies at fixed halo mass (i.e., $\lambda$ and $M_I^{\rm CG}$ are \emph{anti-correlated} at fixed halo mass).  \vspace{0.5\baselineskip}

\item In our second model (MOD2), we reassign all centrals galaxies such that
  heavier halos contain fainter centrals at fixed richness (i.e., $M_{200b}$ and $M_I^{\rm CG}$ are maximally anti-correlated at fixed $\lambda$). In this
  model, clusters with lower values of $\lambda$ have fainter
  central galaxies at fixed halo mass (i.e., $\lambda$ and $M_I^{\rm CG}$ are \emph{correlated} at fixed halo mass). 
\end{enumerate}

\begin{figure*}
\centering
\includegraphics[width=16cm,clip]{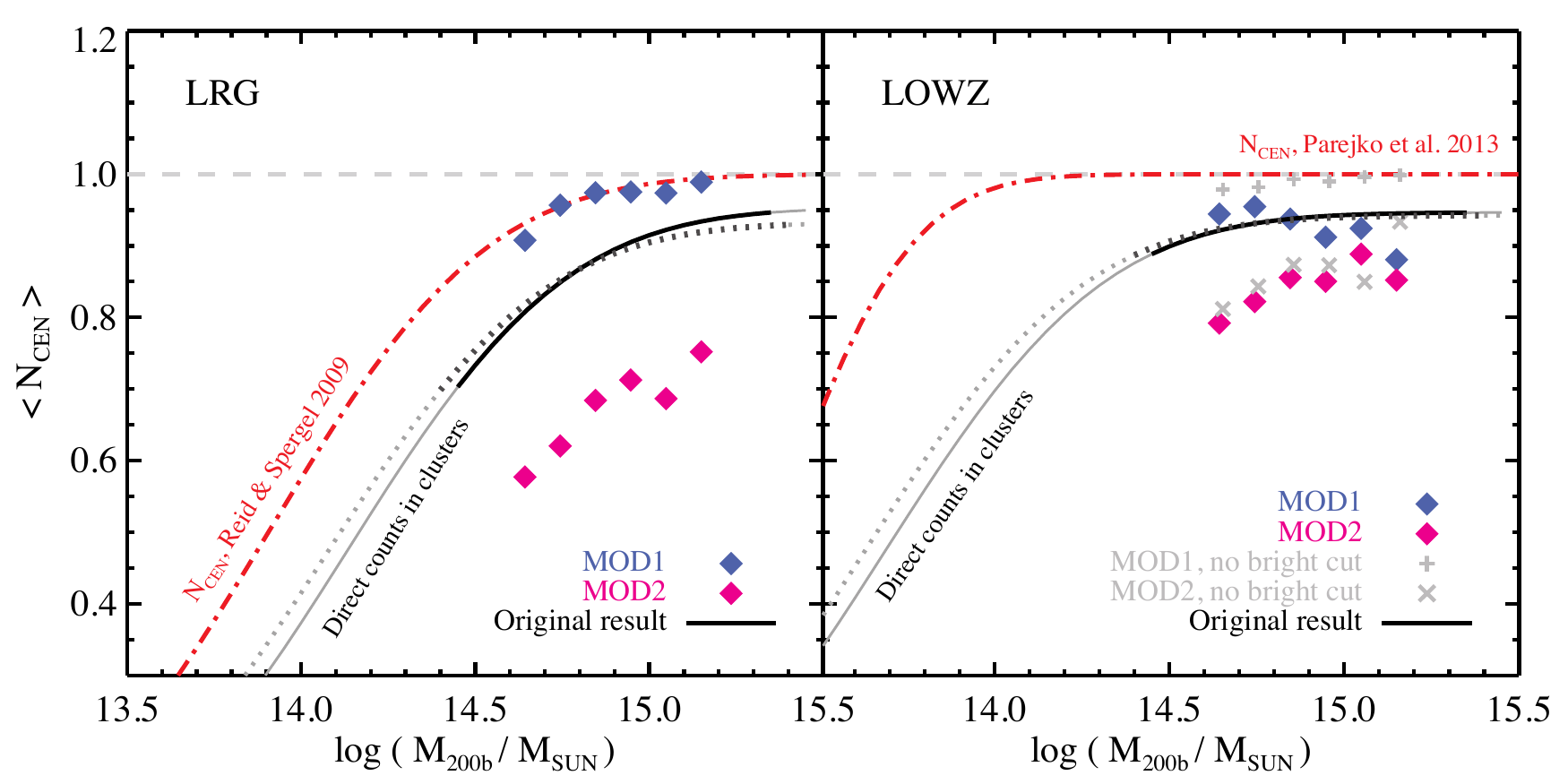}
\caption{Tests to explain the origin of the $N_\text{cen} (M_{200b})$
  discrepancy. Our original results from Figure \ref{fig:hod} are
  shown with black/grey solid curves. First, we test the impact of the
  mass-richness relation. The dotted grey lines shows $N_\text{cen}
  (M_{200b})$ inferred using an updated mass-richness relation with a
  slope of 1.2, instead of 1.08, in Equation
  \eqref{eq:p-m200-lambda}. The updated mass-richness relation only
  has a minor impact on our results. Second, we test if correlations
  between halo mass, richness, and magnitude of the central galaxy may
  impact our results. In both panels, the blue data points show MOD1,
  in which halo mass and central galaxy luminosity are correlated at
  fixed richness (or brighter centrals live in less rich clusters at
  fixed halo mass). The magenta points show MOD2, in which halo mass
  and central galaxy luminosity are anti-correlated at fixed halo
  mass. In the case of MOD1, the bright LOWZ cut removes the brightest
  centrals which are in the most massive halos causing $N_\text{cen}
  (M_{200b})$ to decrease in higher mass halos. Grey crosses show MOD1
  when the bright cut is not applied -- demonstrating that the
  down-turn in indeed caused by the bright cut. The red curves in the
  left and right panels show clustering results from
  \citet{Reid:2009} and \citet{Parejko:2013}, respectively. }
\label{fig:ncen-test}
\end{figure*}

We compute $N_\text{cen}({M_{200b}})$ for MOD1 and MOD2 and the
results are shown in Figure \ref{fig:ncen-test}. Interestingly, we
find that correlations between halo mass, richness, and central galaxy
luminosity may have a {\em large} impact on our inferences about
$N_\text{cen} (M_{200b})$. If the magnitude of the central galaxy is
maximally correlated with halo mass at fixed richness, then our
results would agree with those of \citet{Reid:2009}. For LOWZ
galaxies, however, there is a conflicting effect due to the bright cut
given by Equation \eqref{eq:lowz-eq3}. This cut removes bright galaxies,
which, in MOD1, are preferentially located in the most massive
halos. Therefore, $N_\text{cen}({M_{200b}})$ {\em decreases} at high
halo masses for MOD1. By removing the bright cut, we obtain similar 
results for the LOWZ and classical LRG samples. However, a fair comparison to 
\citet{Parejko:2013} requires the bright cut for LOWZ selection.

A strong anti-correlation between central luminosity and cluster
richness (MOD1) at fixed halo mass would bring our results into good
agreement with \citet{Reid:2009} and $N_\text{cen}({M_{200b}})$ would
converge to unity at large halo masses. It is possible that merging
might lead to such an anti-correlation. When a satellite merges with
the central galaxy, the richness of the cluster decreases while the
magnitude of the central galaxy increases. In this scenario, halo mass
estimators based on combinations of $M_I^{\rm CG}$ and $\lambda$ would
result in a smaller scatter than halo mass estimators based on just
$\lambda$. In fact, \citet{Reyes:2008aa} present evidence this may be
the case using the MaxBCG cluster catalog.

\begin{figure}
\centering
\includegraphics[width=8cm,clip]{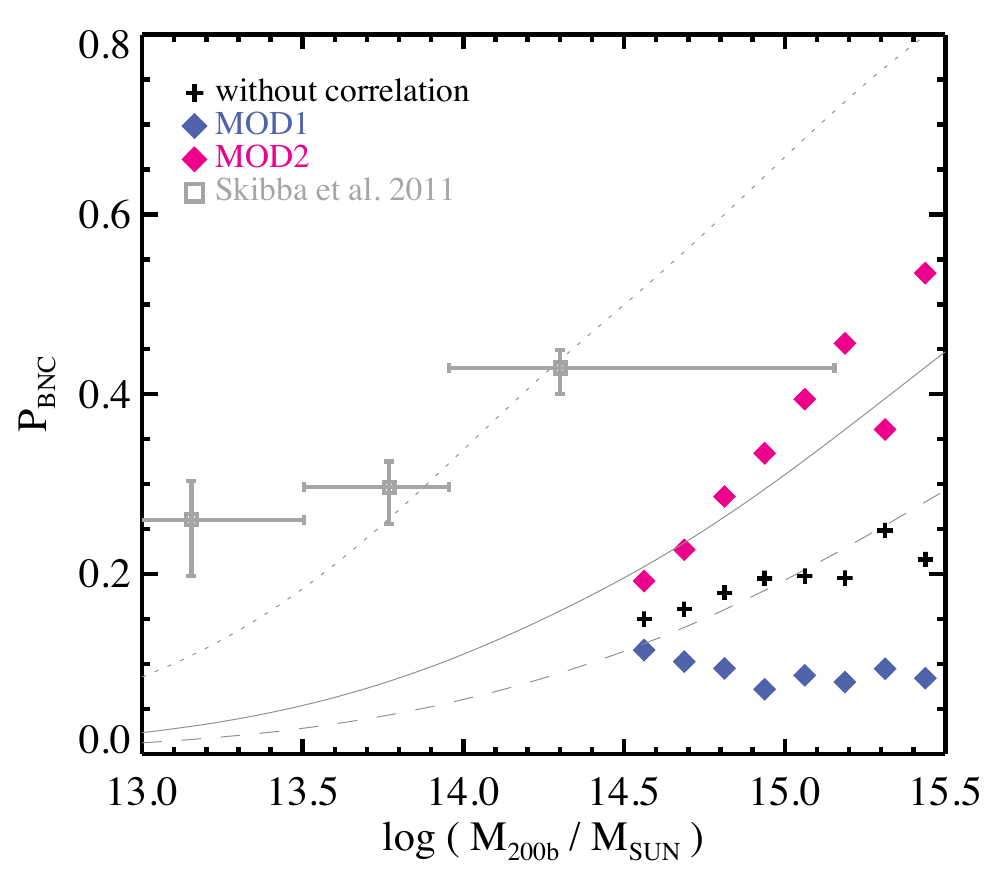}
\caption{Impact of MOD1 and MOD2 on our inference of $P_\text{BNC}$.
  Black plus signs, blue and magenta diamonds represent $P_\text{BNC}$ 
  values when there is no correlation between central luminosity and 
  cluster richness, and cases of MOD1 and MOD2, respectively.
  Correlations between central luminosity and cluster
  richness clearly have a large impact on $P_\text{BNC}$. The MOD1
  model brings our results into closer agreement with \citet{Reid:2009}
  but leads to a larger tension with \citet{Skibba:2011}.}
\label{fig:pbnc_mock}
\end{figure}

Figure \ref{fig:ncen-test} demonstrates that correlations between
central luminosity and cluster richness would clearly impact our
inference about $N_\text{cen}({M_{200b}})$. Figure \ref{fig:pbnc_mock}
also demonstrates that such correlations would also impact our
inference of $P_\text{BNC}$. To match the \citet{Reid:2009} result, we
must assume that central galaxy luminosity and cluster richness are
strongly anti-correlated. While such a strong anti-correlation seems
unlikely, preliminary results from Reddick et al. in prep based on an
analysis of the Conditional Luminosity Function (CLF) of \redmapper{}
cluster members also suggest a strong anti-correlation between
richness and central galaxy luminosity.

In conclusion, a strong anti-correlation between central luminosity
and cluster richness at fixed halo mass is required in order to
reconcile our results with those based on clustering studies. However,
further investigation is needed to confirm or repudiate this
hypothesis. An interesting direction for future work will be follow-up
on our study by performing a {\em joint} CLF analysis of LRG
clustering, galaxy-galaxy lensing, and direct counts in clusters
\citep{Tinker:2012}.

\subsection{How Robust is $N_\text{cen}$ Inferred from Clustering and CIC Studies?}

Finally, we note that the values of $N_\text{cen}$ from studies
similar to \citet{Reid:2009} that use observables such as
counts-in-cylinders or clustering could be dominated by theoretical
prejudices. For example, at the low mass end, HOD models often make
the standard assumption that a halo cannot host a satellite galaxy if
it does not also host a central galaxy from the same sample. Under
this assumption, most LRGs in low mass halos (which contain on average
only one LRG or less) will be centrals. But given that observations of
counts-in-cylinders probe $N_\text{tot}$ they should not be able
distinguish between this standard scenario and a scenario in which a
fraction of LRGs in low mass halos are in fact satellites.

The clustering signal may also be relatively insensitive to such
effects. In these small halos there are no central-satellite or
satellite-satellite pairs of galaxies. Therefore, the difference in
the clustering between the two scenarios will arise only due to galaxy
pairs in 2 different halos (the 2-halo term), which also may not be
heavily affected. In particular, if non-central LRGs in low mass
halos follow a radial distribution peaked towards the center as shown
in Figure~\ref{fig:distance-offset}, the differences in the clustering
signal may be small.

Similar arguments hold at the high mass end. HOD models often assume
that $N_\text{cen}$ approaches unity at large halo masses. At these
mass scales, the one halo term in the clustering signal is dominated
by the satellite-satellite term (which is not sensitive to
$N_\text{cen}$), except possibly at the smallest scales where the
observational uncertainties are significantly larger
\citep{Masjedi:2006aa}. As argued before, counts-in-cylinders will not
be able to make a distinction between centrals and satellites either.

The effect of the above theoretical prejudices on the $N_\text{cen}$
constraints obtained from studies relying on counts-in-cylinders or
clustering can be studied in detail using mock galaxy catalogs. Such
an investigation, however, is beyond the scope of the current paper,
and is left for future work.

\section{Summary and Conclusions}
\label{sec:conclusions}

In this paper, we investigate the connection between central galaxies
in clusters and two classes of red luminous spectroscopic galaxies:
``classical LRGs'', which were targeted as part of the SDSS-I and
SDSS-II programs \citep{Eisenstein:2001}, and LOWZ galaxies from
the SDSS-III BOSS program \citep{Dawson:2013}. The aim of this paper
is to address two different questions: (i) what fraction of cluster
central galaxies are classified as either an LRG or a LOWZ galaxy, and
(ii) when a central cluster galaxy is either a LRG or a LOWZ galaxy, how
often is the central galaxy also the brightest LRG in the cluster.

To tackle these two questions, we use the state-of-the art
\redmapper{} cluster catalog in the redshift range $0.16<z<0.33$ and
in the richness range $\lambda>20$, corresponding to a halo masses
above $2.44 \times 10^{14} M_\odot$. To avoid complications due to fiber collisions,
we construct a set of photometric LRG/LOWZ cluster members. Extensive
tests show that these photometric samples are in excellent agreement with
the original LRG and LOWZ target catalogs ($>95\%$, see Appendix).

A key feature of the \redmapper{} cluster catalog is that it defines
cluster members and central galaxies probabilistically. Using the
centering probabilities given by redMaPPer, we derive the central
occupation of LRGs and LOWZs as a function of richness, $N_\text{cen}
(\lambda)$, and halo mass, $N_\text{cen} (M_{200b})$. We find the
striking result that the central occupation function for both classical LRGs and
LOWZ galaxies does not converge to unity, even for large clusters. Instead,
$N_\text{cen} (\lambda)$ converges to $\sim0.95$.  This can be
explained by the large scatter in central galaxy luminosity at fixed
richness and the LRG/LOWZ selection cuts, which both impose absolute
magnitude limits that exclude central galaxies in the \redmapper{}
catalog. 

When we naively convert $N_\text{cen}(\lambda)$ into $N_\text{cen}
(M_{200b})$, assuming that the halo mass only depends on cluster
richness, the central occupation still only converges to  $0.95$ for
massive halos. This is in conflict with the assumptions behind most
HOD-type studies of the two-point correlation function.

While inaccuracies in the \redmapper{} centroids could cause this
discrepancy, we will show in Paper II that the \redmapper{} centering
probabilities are reliable. Instead, we show that if the magnitude of
the central galaxy is maximally correlated with halo mass at fixed
richness (i.e., $\lambda$ and $M_I^{\rm CG}$ are
\emph{anti-correlated} at fixed halo mass), then our results for
``classical'' LRGs would agree with those of \citet{Reid:2009}. This
maximal anti-correlation does not alleviate the tension between our
results for LOWZ galaxies and those of \citet{Parejko:2013}. This is
because the LOWZ selection includes a bright cut, and, by correlating
galaxy brightness and halo mass, we systematically lower the fraction
of central LOWZ galaxies in massive halos, thus decreasing
$N_\text{cen}(M_{200b})$ for massive clusters.

Our results may indicate a strong correlation between central galaxy
luminosity and halo mass at fixed cluster richness. This is equivalent
to a strong \emph{anti-correlation} between central galaxy brightness
and richness at fixed halo mass. It is possible that merging might
lead to this anti-correlation: mergers between satellite galaxies and
the central galaxy will increase the central galaxy luminosity and
decrease the cluster richness. In fact, examples of
  the extreme cases of such a correlation have been well studied and
  are known as ``fossil groups''
  \citep[e.g,][]{Jones:2003,DOnghia:2005,Zentner:2005}. However,
further investigation will be necessary in order to confirm if this
correlation may be strong enough to fully explain our results. Another
possibility is that the HOD models from counts-in-cylinders or
clustering may be subject to theoretical prejudices in this high halo
mass regime. It is possible that the differences that we observe for
the central occupation function may be caused by a combination of both
effects. An interesting direction for future work will be to try to
disentangle these effects by performing a {\em joint} CLF analysis of
LRG clustering, galaxy-galaxy lensing, and direct counts in clusters
as in \citet{Tinker:2012}.

We also investigate how often the brightest cluster member is not the
central galaxy, $P_\text{BNC}$. Using both the cluster member
probabilities and the central galaxy probabilities from \redmapper{},
we find that in $20 \sim 30\%$ of clusters the brightest galaxy is not
the central galaxy. Our estimate is in good agreement with a CLF
analysis by Reddick et al. in prep, but our measurement of
$P_\text{BNC}$ is nearly a factor of two lower than that of
\citet{Skibba:2011}. The large discrepancy with \citet{Skibba:2011}
may be in part due to improvements in the SDSS sky subtraction between
DR4 \citep[used by][]{Skibba:2011} and DR8 (used here). Reddick et
al. in prep demonstrate that improved sky subtraction around luminous,
extended galaxies tends to increase the difference in luminosity
between central and satellite galaxies, thus substantially decreasing
$P_\text{BNC}$. However, further testing is required to determine if
the differences in photometry apply to galaxies in the redshift range
studied here. We have also quantified the radial distribution of
bright non-central LRGs and found that they follow a substantially
different distribution (truncated at the outskirts) compared to the
distribution for all cluster members.

Because we find a significant fraction of clusters in which the BLRG
is not the central galaxy, using the brightest galaxy as a tracer for
the cluster center will lead to errors due to ``mis-centering,''
particularly in RSD studies based on LRG ``halo catalogs''
\citep[e.g.][]{Reid:2010, Hikage:2013}. The analysis in
\citet{Hikage:2013} suggests a mis-centering fraction for BLRGs of $30
\sim 40\%$, in rough agreement with the $20 \sim 30\%$ mis-centering
we find above. \citet{Hikage:2013} demonstrate that mis-centering of
this magnitude may significantly effect cosmological parameter
estimation in future surveys, such as eBOSS, PFS, and DESI. Future
work will investigate how well the BLRG satellite population can be
characterized using a joint analysis of BLRG clustering and
information from cluster catalogues, such as \redmapper{}. By
characterizing satellite BLRGs, the effects of mis-centering may be
mitigated.

The results presented here rely on the assumption that the probability
values from redMaPPer, particularly those for the cluster center, are
accurate. A comparison of X-ray to optical centers in galaxy clusters
with available data suggests that the \redmapper{} centering
probabilities are indeed accurate \citep{Rozo:2014}. However, the
sample sizes for these comparisons are still small, and rely heavily
on X-ray selected sub-samples of clusters.  More detailed
investigation into the \redmapper{} centering probabilities is
warranted. Following the methods of \citet{George:2012} and
\citet{Hikage:2013}, in Paper II, we will perform tests of the
\redmapper{} centering probabilities by using a combination of weak
gravitational lensing and the projected correlation between the
LRG-inferred halos and a fainter photometric sample of
galaxies. Preliminary results from Paper II indicate that the
redMaPPer centroids are better tracers of halo centers than the
brightest cluster members. This lends credence to both the
\redmapper{} centering probabilities, and the results presented in
this work.

In conclusion, LRGs are considered among the best understood samples
of galaxies and are expected to occupy halos in a relatively simple
way. However, in this paper we have shown that even these relatively
simple galaxies still harbor suprises and that the connection between
LRGs and dark matter halos may not be straightforward.

\section*{Acknowledgments}

We thank Tamas Budavari for helping us with querying the SDSS
database. Jeremy Tinker kindly provided us with the LOWZ target
catalog, and Naoshi Sugiyama supported us with helpful discussions. We
thank the anonymous referee for useful comments that helped improve
this paper. AL, CL, AM, SM, SS and BV were supported by World Premier
International Research Center Initiative (WPI Initiative), MEXT,
Japan. RM acknowledges the support of the Department of Energy Early
Career Award program. BV was supported by the Kakenhi Grant-in-Aid for
Young Scientists (B)(26870140) from the Japan Society for the
Promotion of Science (JSPS). AM was supported by a research fellowship
from the Japan Society for the Promotion of Science (JSPS).


\bibliographystyle{mn2e}
\bibliography{references}
\label{lastpage}

\appendix
\section{Our photometric sample and SDSS target selection}

Our LRG sample selection is described in Section \ref{sec:lrg}. Here, we compare this sample to the actual LRG sample targeted for spectroscopy in SDSS \citep{Eisenstein:2001}. Although the cuts described above are the same as those for the targeted sample, the SDSS photometry changed slightly between DR7 and DR8. As we show below, this slightly alters our LRG sample when compared to the targeted sample. To do this comparison, we introduce the following new samples:
\begin{enumerate}
   \item DR7PhotLRG: Photometrically selected LRGs with DR7 photometry
   \item TargLRG: Original set of LRGs that are targeted by SDSS
   \item TargLOWZ: Original set of LOWZs that are targeted by SDSS
\end{enumerate}

\subsection{DR8PhotLRG, DR7PhotLRG and TargLRG}
The SDSS LRGs selected for spectroscopy are flagged
\verb+TARGET_GALAXY_RED+, and we queried DR7 to get all such
galaxies. We call these SDSS LRGs ``TargLRGs". These galaxies are
selected using the cuts described in Section \ref{sec:classical-LRGs},
but using earlier reductions of the SDSS data.

We note here that there is an implicit bright cut in TargLRG, due to
the bright cut in the SDSS main sample, which is not part of our
photometric LRG selection. The bright limit is included because very
bright objects may cause contamination of spectra or saturation of the
CCDs. Galaxies with apparent magnitudes brighter than 15 in $g$- or
$r$-band or 14.5 in $i$-band, or Petrosian magnitudes brighter than
$r_\text{Petro} < 15$ are excluded from SDSS main sample, and,
therefore from TargLRG. However, since every galaxy in our redMaPPer
cluster catalog is fainter than this limit, the bright limit for
classical LRGs is not relevant to our work.

We compare TargLRG to our sample of photometrically selected LRGs,
DR8PhotLRG. Since the DR8 sky coverage is wider than that of DR7, we
do the comparison in a region where DR8 and DR7 overlap, i.e., the
north Galactic cap. However, we find that the TargLRG sample contains
holes in this region, due to missing runs. Therefore, we restrict our
comparison to smaller region in the north Galactic cap: $120 <
\mathrm{ra} < 250$ and $20 < \mathrm{dec} < 50$. In this region, there
are 3,549 DR8PhotLRGs and 3,542 TargLRGs; 3,191 galaxies are in both
samples, which is 89.9\% of the DR8PhotLRG sample and 90.1\% of
TargLRG sample.

We perform the same comparison, replacing DR8PhotLRGs with
DR7PhotLRGs. These samples are selected identically, but the latter
uses DR7 photometry. Since the LRG target catalog was made before DR8,
and there are small changes in the photometry between DR8 and earlier
data releases, it is worthwhile to compare TargLRGs with
DR7PhotLRGs. Here we again take the region limited by $120 <
\mathrm{ra} < 250$ and $20 < \mathrm{dec} < 50$ in the north Galactic
cap so as not to be affected by missing runs in TargLRG. In this
region, there are 3532 DR7PhotLRGs and 3542 TargLRGs. The samples
overlap for 3323 galaxies, which is 97.9\% of DR7PhotLRG and 97.7\% of
TargLRG.

In conclusion, when we use DR7 photometry, we find a 6\% difference
between the original LRG target catalog and our photometric
selection. When we use DR8 photometry, the difference increases to
12\%. This demonstrates, that while similar, our classical LRG sample
is not identical to the SDSS spectroscopic LRG sample and care must be
taken when comparing the two. The source of these differences is
likely scatter in the photometry between reductions. Directly
comparing DR7 and DR8 photometry for the same galaxies, we find the
RMS scatter in the magnitudes and colors is $\sim 0.07$ mags, with no
systematic offsets. Photometric scatter will affect faint LRGs more
than bright LRGs. Reassuringly, when comparing TargLRG, DR7PhotLRG,
and DR8PhotLRG, galaxies in one sample but not the others are on
average half a magnitude fainter than typical LRGs.  Photometric
scatter mainly affects the LRG sample near the color-magnitude cuts
for LRG selection, but the overall number of LRGs selected in each
sample is nearly unchanged. Therefore, while the samples of LRGs
cannot be directly compared, they are equivalent for our analysis.

\subsection{BOSS LOWZ target selection}

As with classical LRGs, we compare our original PhotLOWZs to the LOWZ
target catalog. For the LOWZ target catalog, we use
``bosstarget-lrg-main007-collate.fits''
\footnote{http://www.sdss3.org/dr10/algorithms/boss\_target\_selection.php}. This
catalog contains LOWZ galaxies, as well as CMASS-selected galaxies at
higher redshifts ($0.4<z<0.7$). We select LOWZ galaxies with the flag
\begin{center}
(boss\_target1 AND $2^0$) NE 0.
\end{center}
We call the LOWZs in the target catalog ``TargLOWZs". Before comparing
TargLOWZs and PhotLOWZs, we need to remove regions with problems in
the target catalog. Due to a bug in the star-galaxy separation used to
create the target catalog, there are some regions we need to
exclude. To remove undesired regions, we remove objects from TargLOWZ
with
    \[ \mathrm{TILE} < 10324. \]
Then we simply take the regions that are not affected by this bug. They are:
\begin{itemize}
\item $0<\mathrm{RA}<50$, $2<\mathrm{DEC}<40$
\item $140<\mathrm{RA}<270$, $45<\mathrm{DEC}<70$
\item $100<\mathrm{RA}<220$, $6<\mathrm{DEC}<28$
\item $310<\mathrm{RA}<360$, $-15<\mathrm{DEC}<40$
\end{itemize}

There are 12,778 PhotLOWZs in these regions, 12,294 of which (96.2\%)
are also selected as TargLOWZ. As with classical LRGs, the photometric
LOWZ selection closely mimics the actual spectroscopic LOWZ
sample. The small differences can be explained by small changes in the
photometry used for target selection over the course of the BOSS
survey.

\section{SDSS queries}
\subsection{How to get photometric properties from DR8}

To obtain data from SDSS, we used the SDSS CasJobs tool \footnote{http://skyserver.sdss3.org/CasJobs/}. First, we match objects from the redMaPPer catalog to SDSS data in RA and Dec using a 1.5 arcsecond match radius. The results of this match are stored in \verb+matchtable+. We select the properties of the matched redMaPPer galaxies from SDSS DR8 using the \verb+SQL+ query: 

\vspace{0.5\baselineskip}

\noindent
{\tt SELECT \\
  matchtable.search\_id, 
  matchtable.ra, 
  matchtable.dec, 
  p.modelMag\_r, 
  p.modelmag\_i, 
  p.modelmag\_g, 
  p.modelmag\_z, 
  p.modelmag\_u, 
  p.modelmagerr\_u, 
  p.modelmagerr\_g, 
  p.modelmagerr\_r, 
  p.modelmagerr\_i, 
  p.modelmagerr\_z, 
  p.cmodelMag\_r, 
  p.cmodelmag\_i, 
  p.cmodelmag\_g, 
  p.cmodelmag\_z, 
  p.cmodelmag\_u, 
  p.cmodelmagerr\_u, 
  p.cmodelmagerr\_g, 
  p.cmodelmagerr\_r, 
  p.cmodelmagerr\_i, 
  p.cmodelmagerr\_z, 
  p.extinction\_r, 
  p.extinction\_i, 
  p.extinction\_g, 
  p.extinction\_z, 
  p.extinction\_u, 
  p.petromag\_r, 
  p.psfmag\_r, 
  p.petror50\_r, 
  p.devrad\_i, 
  p.flags, 
  p.objid  \\
INTO mydb.outputtable \\
FROM mydb.matchtable \\
JOIN PhotoPrimary as p on matchtable.matched\_id=p.objid \\
ORDER BY matchtable.search\_id}

\vspace{0.5\baselineskip}

\subsection{How to get actual LRG target from SDSS}
To get the SDSS DR7 LRG target catalog (TargLRGs), we queried the SDSS
DR7 database for all galaxies flagged as \verb+TARGET_GALAXY_RED+:

\vspace{0.5\baselineskip}

\noindent
{\tt SELECT \\
 run,
 camCol,
 rerun,
 field,
 objID,
 ra,
  dec,
  modelmag\_u, modelmag\_g, modelmag\_r, modelmag\_i, modelmag\_z,
  modelmagerr\_u, modelmagerr\_g, modelmagerr\_r, modelmagerr\_i, modelmagerr\_z,
  extinction\_u, extinction\_g, extinction\_r, extinction\_i, extinction\_z,
  petromag\_r, psfmag\_r, petror50\_r, devrad\_i \\
into mydb.DR7\_LRG\_all \\
from TargPhotoPrimary \\
WHERE (primtarget \& \\
  dbo.fPrimTarget('TARGET\_GALAXY\_RED')) != 0}

\vspace{0.5\baselineskip}


\end{document}